\documentclass[sigconf,screen]{acmart}

\usepackage{multicol}
\usepackage{colortbl}
\usepackage{multirow}
\usepackage{hhline}
\usepackage{tabularx}
\usepackage{subcaption} 
\usepackage{alltt}
\usepackage{listings}
\usepackage{cuted}

\usepackage{verbatim} %

\newcommand{\RootFile}{MAIN_PAPER.tex}

\newcommand{\PrintWordcountReport}{%
  \immediate\write18{texcount -inc -merge -sub=section -utf8 "\RootFile" > "\jobname.wc"}%
 \clearpage
 \section*{Word count by section}
  \verbatiminput{\jobname.wc}%
}

\AtBeginDocument{%
  }

\copyrightyear{2026}
\acmYear{2026}
\setcopyright{cc}
\setcctype{by-nc-nd}
\acmConference[CHI '26]{Proceedings of the 2026 CHI Conference on Human Factors in Computing Systems}{April 13--17, 2026}{Barcelona, Spain}
\acmBooktitle{Proceedings of the 2026 CHI Conference on Human Factors in Computing Systems (CHI '26), April 13--17, 2026, Barcelona, Spain}
\acmPrice{}
\acmDOI{10.1145/3772318.3791199}
\acmISBN{979-8-4007-2278-3/2026/04}

\begin{document}

\newcolumntype{L}[1]{>{\raggedright\arraybackslash}p{#1}}

\newcommand{\iquote}[1]{\textit{\color{darkgray}{#1}}}

\newcommand{\pid}[1]{{\fontfamily{cmss}\selectfont{\footnotesize{{\color{darkgray}{#1}}}}}}

\newenvironment{smallquote}
  {\begin{quote}\normalsize}
  {\end{quote}}
  
\newtoggle{comments}
\togglefalse{comments}

\iftoggle{comments} {
\newcommand{\rr}[1]{{\color{red}{#1}\normalfont}}
\newcommand{\sean}[1]{{\color{magenta}{SR: #1}\normalfont}}
\newcommand{\ava}[1]{{\color{cyan}{AS: #1}\normalfont}}
\newcommand{\panda}[1]{{\color{green}{PP: #1}\normalfont}}
}{
  \newcommand {\lev}[1]{}
  \newcommand {\sean}[1]{}
  \newcommand {\ava}[1]{}
  \newcommand {\panda}[1]{}
 }

\title[Nudging Attention to Workplace Meeting Goals]{Nudging Attention to Workplace Meeting Goals: A Large-Scale, Preregistered Field Experiment}

\author{Lev Tankelevitch}
\authornote{Both authors contributed equally to this research.}
\affiliation{%
  \institution{Microsoft Research}
  \city{Cambridge}
  \country{United Kingdom}
}
\email{lev.tankelevitch@microsoft.com}

\author{Ava Elizabeth Scott}
\authornotemark[1]
\authornote{The work was done when the co-author was employed at Microsoft.}
\affiliation{%
  \institution{University of Copenhagen}
  \city{Copenhagen}
  \country{Denmark}
}
\email{Ava.scott@di.ku.dk}

\author{Nagaravind Challakere}
\affiliation{%
 \institution{Microsoft}
 \city{Redmond}
 \state{WA}
 \country{USA}
 \postcode{98052}
}
\email{Nagaravind.Challakere@microsoft.com}

\author{Payod Panda}
\affiliation{%
  \institution{Microsoft Research}
  \city{Cambridge}
  \country{United Kingdom}}
\email{payod.panda@microsoft.com}

\author{Sean Rintel}
\affiliation{%
  \institution{Microsoft Research}
  \city{Cambridge}
  \country{United Kingdom}}
\email{serintel@microsoft.com}

\renewcommand{\shortauthors}{Tankelevitch, Scott, et al.}

\begin{abstract} %

Ineffective meetings are pervasive. Thinking ahead explicitly about meeting goals may improve effectiveness, but current collaboration platforms lack integrated support. We tested a lightweight goal-reflection intervention in a preregistered field experiment in a global technology company (361 employees, 7196 meetings). Over two weeks, workers in the treatment group completed brief pre-meeting surveys in their collaboration platform, nudging attention to goals for upcoming meetings. To measure impact, both treatment and control groups completed post-meeting surveys about meeting effectiveness. While the intervention impact on meeting effectiveness was not statistically significant, mixed‑methods findings revealed improvements in self‑reported awareness and behaviour across both groups, with post‑meeting surveys unintentionally functioning as an intervention. We highlight the promise of supporting goal reflection, while noting challenges of evaluating and supporting workplace reflection for meetings, including workflow and collaboration norms, and attitudes and behaviours around meeting preparation. We conclude with implications for designing technological support for meeting intentionality.

\end{abstract}



\maketitle

\section{Introduction}\label{sec:intro}

Shared understanding of goals should make meetings effective \cite{allenLinkingPremeetingCommunication2014b, rogelbergSurprisingScienceMeetings2019, bang_effectiveness_2010}, yet a persistent disconnect is evident in decades of dissatisfaction with purposeless meetings \cite{microsoft_WTI_2023, cohen_meeting_2011, geimer_meetings_2015, leach_perceived_2009, luong_meetings_2005, nixon_impact_1992, rogelberg_not_2006, rogelberg_thirty_2025}. Bridging this gap depends on deliberate reflection on meeting goals, which can improve meeting preparation awareness and behaviour \cite{bang_effectiveness_2010,scott_mental_2024}. However, such reflection rarely fits easily into fast-paced work. This is a specific instance of the general problem of nudging people towards workplace reflection \cite{meyer_enabling_2021, fleck_reflecting_2010, baumer_reviewing_2014}.  A key challenge is designing interfaces that make reflection a quick, convenient, and impactful part of the meeting lifecycle, balancing the effort of reflection with the demands of everyday meetings and broader work contexts \cite{sweller_element_2010, sweller_cognitive_1988,scott_mental_2024}.
 This paper contributes to design explorations of workplace reflection technologies for thinking ahead, and articulating and communicating meeting goals.%

While there is growing support for the value of workplace reflection \cite{meyer_enabling_2021,williams_supporting_2018,samrose_meetingcoach_2021,kocielnik_designing_2018,scott_what_2025,prilla_supporting_2014,bentvelzen_revisiting_2022}, much of the research is conducted with limited ecological validity, relying on out-of-context reflection, unfamiliar interfaces, or small samples \cite{ilies2024_diarystudy, cho_reflectiveinterfaces_2022, kocielnik_designing_2018}. To address this, we ran a two-week longitudinal field experiment with 361 employees at a global technology firm. Delivered within their workplace collaboration platform, the treatment group received \textit{pre-meeting surveys} prompting reflection on goals for their upcoming meetings. To measure impact, both the treatment and control groups received \textit{post-meeting surveys} asking for ratings of perceived meeting effectiveness. Our aims were to identify a minimal viable intervention and explore contextual issues for designing for meeting intentionality.

The pre-meeting survey intervention did not lead to statistically significant improvements in perceived meeting effectiveness. However, \textit{both treatment and control} groups showed pre-post improvements in self-reported goal clarity and communication---upstream drivers of effective meetings---with mixed-methods findings suggesting that post-meeting surveys unintentionally functioned as an intervention for both groups. As such, while noting evaluation limitations, the study highlights the overall promise of nudging reflection on meeting goals for improving meeting intentionality and effectiveness. We explore the complexities of fostering meeting intentionality, including challenges related to workflow integration and prevailing collaboration norms, attitudes, and behaviours around meeting preparation. Finally, we outline design considerations for technologies that nudge attention to meeting goals within the broader context of workplace reflection. Our contributions are:
\begin{itemize}
    \item Mixed-methods findings from a large-scale longitudinal field experiment testing a workplace meeting goal reflection intervention.
    \item Characterisation of challenges to reflection and meeting intentionality in a global organisation.
    \item Methodological insights on conducting rigorous and ecologically valid evaluations of reflection support at scale.
    \item Design implications for timing, targeting, modality, and impact of meeting goal reflection support within collaborative knowledge workflows.
\end{itemize}

\section{Related Work}\label{sec:relatedwork}

\subsection{Meetings and Purpose}

HCI and CSCW have focused on improving meetings as collaborative experiences, initially focused on the value of video meetings \cite{chapanis_studies_1972,pye&williams_teleconferencing_1977}, turning to the demands of representation and inclusion of people and places \cite{finn_video-mediated_1997,harrison2009,brehalt_asymmetry,panda_hybridge,jensemil_mirrorverse,neumayr_territoriality,cutler_meeting_2021,sarkar_parallelchat,nowak_spatialaudio,castelli_why_2021, rodegheroPleaseTurnYour2021c,chen_meetmap}. Many pain points in the meeting lifecycle have been addressed, from pre-meeting issues to such as scheduling \cite{Brzozowski2006, Cranshaw2017}, issues during meetings such as facilitation and leadership \cite{benke_leadboski,Shamekhi_facilitorBot}, to post-meeting issues such as recaps and summaries for extracting decisions and actions \cite{junuzovicRequirementsRecommendationsEnhanced2008c, kalnikaiteMarkupYouTalk2012f,vega-oliverosThisConversationWill2010b, kumar_meeting_2022, ter_hoeve_what_2022,wang_meetingbridges_2024,asthana_summaries_2025}, as well as the broader problems of videoconferencing fatigue \cite{bailensonNonverbalOverloadTheoretical2021f,doring_videoconference_2022,riedlStressPotentialVideoconferencing2021,bergmann_meeting_2023}. However, most of these efforts have left the upstream issue of workers’ complaints about meetings lacking purpose unaddressed \cite{microsoft_WTI_2023, cohen_meeting_2011, geimer_meetings_2015, leach_perceived_2009, luong_meetings_2005, nixon_impact_1992, rogelberg_not_2006, rogelberg_thirty_2025}. Recent work has highlighted the role of meeting intentionality \cite{scott_mental_2024} and the potential value of reflection  in supporting meeting effectiveness \cite{scott_what_2025, chen_are_2025, vanukuru_designing_2025}, pointing to a need to support workplace reflection.

\subsection{Reflection and Nudging}

Reflection involves consciously re-evaluating thoughts, ideas, and experiences to guide future behaviour \cite{atkins_reflection_1993, fleck_reflecting_2010, scott_what_2025}, including noticing, understanding, and questioning mental processes \cite{fleming_how_2014, hollis_what_2017}. It can occur before, during, or after an activity \cite{schon2017reflective}. \citet{vanukuru_designing_2025} highlight the temporal nature of meetings, where prospection integrates future visions with present actions \cite{szpunar_taxonomy_2014}, and retrospection connects past interactions across projects \cite{sellen_beyond_2010}. 
Workplace reflection extends these ideas to organisational contexts, where competing priorities, time pressure, and social dynamics shape when and how reflection occurs, requiring a balance between individual cognition and collaborative practices \cite{meyer_enabling_2021,fleck_reflecting_2010, baumer_reviewing_2014}.

Reflection intersects with the behavioural science concept of \textit{nudging}: designing choices to predictably change behaviour without restricting options or significantly altering their economic incentives \cite{thaler2008nudge}. Nudging is grounded in dual-process theories of cognition which posit the existence of a fast \textit{System 1} (representing habits, heuristics, biases etc.), and a slow \textit{System 2} (representing reflection, deliberation, controlled reasoning etc.) \cite{evans_dual-process_2013}. The original proposal for nudging emphasises its low cognitive cost to the decision-maker, aligning it with fast \textit{System 1} thinking---e.g., changing default choices is a classic nudge \cite{thaler2008nudge}. The more recent \textit{`nudge plus'} involves incorporating aspects of reflection, i.e., \textit{System 2} thinking, into nudges to both increase people's agency, and augment behaviour change through conscious self-evaluation \cite{Banerjee2024NudgePlus}. Our intervention is grounded in this latter approach.

\subsection{Supporting Reflection on Intentions in the Workplace}

Interfaces can support reflection by nudging people to consider different perspectives of a situation, scaffolding the identification of priorities and the development of plans of actions to achieve specific goals~\cite{zhu_systematic_2025,bentvelzen_revisiting_2022, slovak_reflective_2017, cho_reflectiveinterfaces_2022}. There is extensive evidence that the articulation of clear goals and plans can improve performance, motivation, and coordination \cite{locke_goal-setting_2005, sheeran_interplay_2005, gollwitzer_implementation_2006, zhu_systematic_2025}. %
Below, we outline research into supporting prospective and retrospective reflection.

\citet{meyer_enabling_2021} found that daily reflective goal-setting via surveys improved productivity among 43 software developers, with 80\% reporting positive behavioural change. A goal-oriented conversational agent, reported by \citet{williams_supporting_2018}, supported work ‘detachment’ and ‘reattachment’ across workdays. Evaluated over two weeks with 34 information workers, it reduced after-hours emails and increased next-day self-reported productivity and engagement. For retrospective reflection, \citet{kocielnik_designing_2018} showed that a conversational agent used by 10 participants over three weeks supported daily workplace reflection on goals, fostering awareness, perspective change, and aiding management and performance. \citet{samrose_meetingcoach_2021} designed an interactive dashboard prompting 23 participants to reflect on behaviours related to meeting effectiveness and inclusiveness, improving post-meeting awareness of dynamics. 

Specifically targeting prospective reflection on meeting goals and challenges, \citet{scott_what_2025} evaluated a generative AI assistant with 18 participants. The assistant facilitated reflective conversations about upcoming meetings, prompting some participants to reconsider their approach and take additional preparatory actions. However, the interaction was perceived as time-consuming and effortful, with participants indicating they would reserve such detailed reflection for important or uncertain meetings. The study also identified barriers to goal reflection, including reluctance to articulate specific goals for predominantly social meetings.

While the studies above demonstrate the value of workplace reflection, they also show that there are substantial research challenges in establishing ecological validity for the practice \cite{bentvelzen_revisiting_2022}. To nudge and capture workplace reflection is not easy, often relying on atypical contexts, whether by requiring it outside normal work routines (e.g., \cite{ilies2024_diarystudy,meyer_enabling_2021}), introducing unfamiliar technologies (e.g., \cite{cho_reflectiveinterfaces_2022,williams_supporting_2018,samrose_meetingcoach_2021,scott_what_2025, prilla_supporting_2014}), or relying on small sample sizes (e.g., \cite{kocielnik_designing_2018,meyer_enabling_2021}). These constraints make it difficult to capture the complexity of real-world work environments, where reflection is shaped by time pressure, competing priorities, and social dynamics \cite{meyer_enabling_2021, fleck_reflecting_2010}. In sum, workplace reflection may benefit from quick, in-context support. Our intervention examines taxonomies of meeting purpose to help achieve this. 

\subsection{Taxonomies of Meeting Purpose}\label{sec:taxonomies}

Previous research into the purpose of meetings has produced various taxonomies \cite{pye_description_1978, romano_meeting_2001, a_allen_understanding_2014, soria_recurring_2022, standaert_empirical_2016, standaert_how_2021, scott_mental_2024, monge_profile_1989, lopez-fresno_what_2022}. For example, \citet{monge_profile_1989} distinguished meetings for conflict resolution, decision-making, and problem solving; \citet{a_allen_understanding_2014} identified content- and action-focused meetings with sub-types; and \citet{lopez-fresno_what_2022} found participants typically viewed meeting purposes instrumentally (e.g., to decide, plan, coordinate) rather than socially (e.g., to persuade, motivate, socialise). \citet{scott_mental_2024} showed workers’ mental models broadly fall into meetings as means to an end or ends in themselves. 

Meeting Science taxonomies reveal issues such as goal clarity shaping roles and effectiveness \cite{geimer_meetings_2015, bang_effectiveness_2010}, mismatches between stated and perceived purposes \cite{lopez-fresno_what_2022}, organiser–attendee perspective gaps \cite{a_allen_understanding_2014}, and conflicts between personal and collective goals \cite{romano_meeting_2001, lopez-fresno_what_2022}. Meeting Science has examined agendas \cite{rogelberg_thirty_2025}, but neither goals nor agenda research includes empirical interventions linking goal awareness to the behaviours that shape agendas and shared understanding for effective meetings. This represents a clear point of disconnect: meetings need purpose, but reflection on purpose is missing. \citet{rogelberg_thirty_2025} noted weaknesses in taxonomies, including limited theoretical grounding, unclear category boundaries, and uncertain predictive value for meeting practices and outcomes. However, no research has examined whether these taxonomies are useful for workers' reflection on meetings.

\section{Research Questions}\label{sec:researchquestions}
The research above demonstrates value in supporting in-depth workplace reflection but complexity in enabling this within normal workflows. There is also a lack of research measuring the impacts of prospective reflection, at a large scale, compared against a control condition. To address these gaps, we ran a field experiment embedding reflection prompts into collaboration software, drawing on the above taxonomies to keep reflection quick and convenient. Our research questions are as follows:
\begin{enumerate}
    \item[\textbf{RQ1.}] Does asking workers to reflect on their goals for upcoming workplace meetings impact meeting effectiveness and engagement and upstream aspects like self-awareness and behaviour change?
    \item[\textbf{RQ2.}] What are workers' experiences of goal-oriented reflection for meetings, attitudes towards the intervention, and barriers to reflection and meeting intentionality?
    \item[\textbf{RQ3.}] What are the implications for designing technological support for meeting reflection and intentionality? 
\end{enumerate}

For RQ1, we hypothesised a directional effect based on prior theory (see §\ref{sec:relatedwork}), but this was not preregistered and should therefore be treated as exploratory:
    \begin{quote}
            \textbf{H1:} Pre-meeting surveys nudging reflection on meeting goals increase self-reported meeting effectiveness and engagement, as well as improve the upstream factors of clarity and communication of goals.
    \end{quote}

\section{Methods}\label{sec:methods}

We investigated how nudging workers to reflect on their goals for upcoming workplace meetings affects self-reported meeting effectiveness and related measures. Participants (n = 361) from a global technology company installed an application into their workplace collaboration platform that randomly assigned them to one of two conditions. In the \textbf{treatment} condition, participants received both pre-meeting surveys (prompting goal reflection) and post-meeting surveys (measuring meeting effectiveness). In the \textbf{control} condition, they received only post-meeting surveys. The application was therefore intended both to deliver the intervention and measure its impact. We describe the details of both aspects in §\ref{subsec:meetingapp} below.

Participants completed meeting surveys for as many workplace meetings as possible over a two-week period. This duration balanced giving participants time to adjust, collecting sufficient data despite incomplete survey responses, and avoiding discouragement from perceived overcommitment. The intervention's longitudinal design also aimed to encourage wider behaviour change beyond any individual meeting targeted for intervention. 

Participants also completed an onboarding survey at the study start, and a debriefing survey after the two weeks. We report quantitative analyses evaluating the impact of the intervention on meeting effectiveness and associated outcomes, as well as mixed-methods analyses of survey data covering participants' experiences in the study, potential mechanisms of any changes, and barriers and contextual factors around the intervention.

We preregistered the study and main analyses on AsPredicted (\#194719). Unless noted otherwise, all analyses follow the preregistration. 

\subsection{Meeting Survey application}\label{subsec:meetingapp}

We built an application to deliver pre- and post-meeting surveys targeting specific meetings, integrated into participants’ collaboration workflows, and timed for delivery before and after meetings. Functionality was enabled by connecting to participants’ calendars and embedding within their primary workplace collaboration software.

\subsubsection{Meeting inclusion criteria} \label{subsubsec:inclusioncriteria}
The application targeted calendar events meeting the following criteria: 
\begin{itemize}
    \item Identified as videoconferencing calls via calendar telemetry to obtain basic metadata 
    \item Scheduled for under 3 hours, assuming longer events were likely workshops or holds
    \item Accepted by the participant if not the organiser
    \item Included at least 2 attendees, including the organiser
    \item Had at least 10 minutes of free time beforehand to allow treatment participants to complete the pre-meeting survey (i.e., we only included the first meeting in a block of back-to-back meetings; to ensure we included sufficient meetings, we did not analogously exclude based on post-meeting availability.)
\end{itemize}

\subsubsection{Intervention surface: Microsoft Teams chat} 
The intervention was designed to work in Microsoft Teams. To maximise post-meeting survey responses, a bot initiated individual chat threads separate from meeting chats but still within Teams chats. %
While organisers often reflect during scheduling, attendees typically accept meetings with minimal engagement, making chat a more viable medium. To reduce technical and trial complexity, chat was chosen as the unified intervention surface.

\subsubsection{Intervention and measurement timing} Pre-meeting surveys were sent 5 minutes before the scheduled start time. As participants both prepare and share pre-reads for meetings at widely variable times in advance (or not at all \cite{scott_mental_2024}), this proximity to the meeting aimed to ensure that participants had sufficient information to enable meaningful reflection on goals. Moreover, given participants' busy schedules, it also aimed to ensure that participants would be able to dedicate a brief moment to reflect \cite{gilbert_strategic_2015, gilbert_outsourcing_2023}, particularly when combined with our meeting inclusion criteria of at least 10 minutes of free time before a meeting (§\ref{subsubsec:inclusioncriteria}). Post-meeting surveys were sent 2 minutes after the scheduled end time, aiming to capture fresh recollections. Participant feedback from an initial pilot (n = 41) did not surface any major issues with these timings (but see §\ref{subsubsec:relativetiming}). %

\subsubsection{Intervention: Pre-meeting surveys for goal reflection}\label{subsubsec:presurveydesign}

\begin{figure*}[h]
\centering
\includegraphics[width=\textwidth]{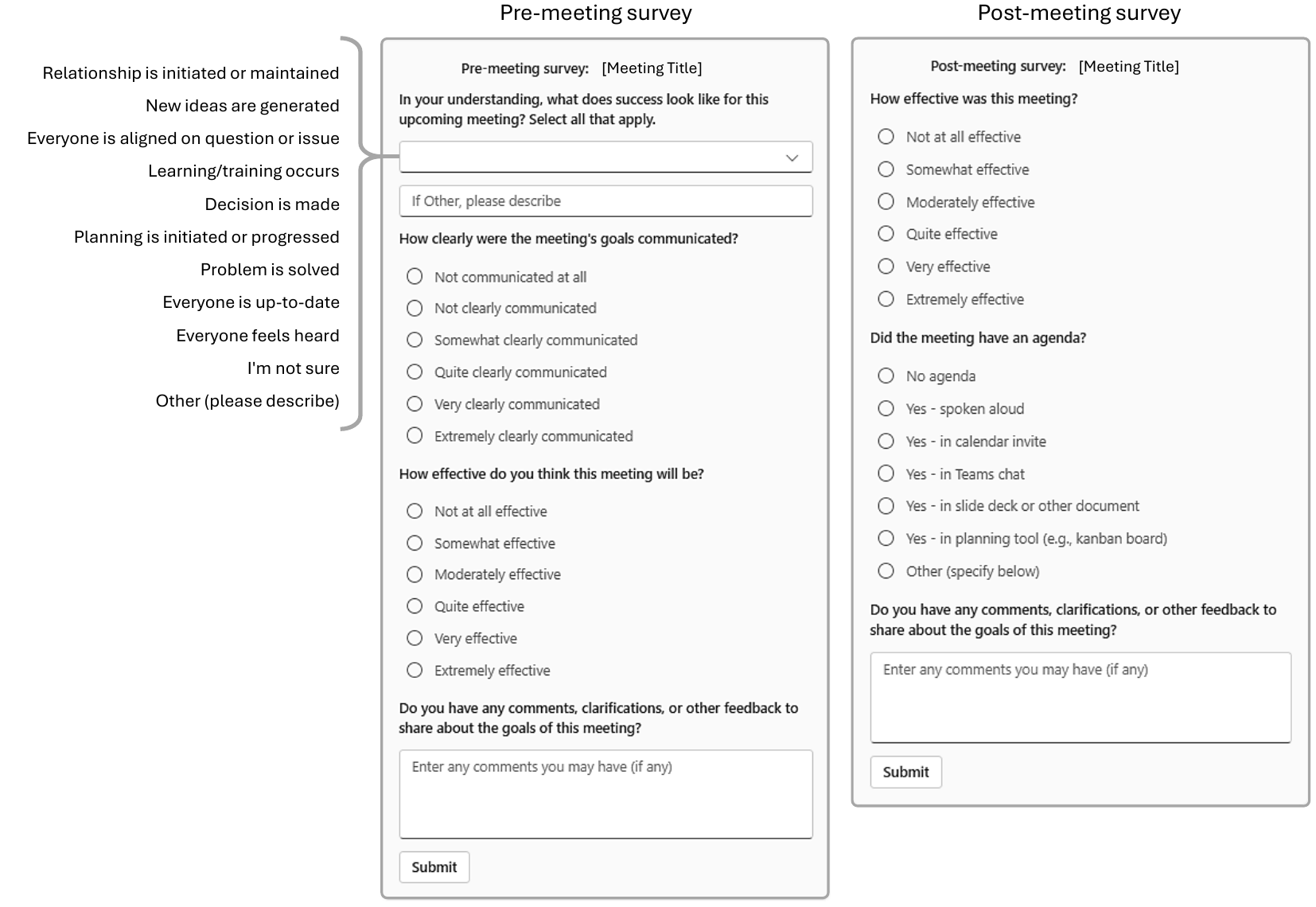}
  \caption{Pre- and post- meeting surveys as they appeared in chat threads via our Meeting Survey application. In the pre-meeting survey, treatment group participants could select multiple goals from 11 options, as well as write in their own goals.}
  \Description{Pre- and post-meeting surveys used to capture goal clarity and meeting effectiveness. The figure shows two side-by-side survey forms designed to assess meetings before and after they occur, illustrating how goal reflection is prompted and later evaluated. On the left, the pre-meeting survey asks participants to define what success looks like for the upcoming meeting by selecting from goal categories (e.g., generating ideas, making decisions, alignment, learning, problem solving, relationship building), optionally describing other goals. It then asks how clearly the meeting’s goals were communicated, how effective the meeting is expected to be, and provides a free-text field for comments about the goals. On the right, the post-meeting survey asks participants to rate how effective the meeting actually was using a Likert scale, whether the meeting had an agenda (and in what form), and to provide any comments or clarifications about the meeting’s goals.}
  \label{fig:surveys}
\end{figure*}

The pre-meeting surveys asked treatment participants to select from a list of 10 goals for their upcoming meeting, with support for custom goal entry (left panel in \autoref{fig:surveys}). Multiple goals could be selected. Instead of open-ended questions typical of reflection interventions \cite{bentvelzen_revisiting_2022, slovak_reflective_2017}, we used a closed format for quick completion %
, cuing recognition and explicit choice without memory recall \cite{norman_design_2013, norman_affordance_1999, plessis_recognition_1994}. The question and response options were phrased as success criteria: ``\textit{In your understanding, what does success look like for this upcoming meeting? Select all that apply}'', with examples such as `\textit{Decision is made}' and `\textit{New ideas are generated}'. A pilot study (n = 41) compared this phrasing to action-oriented (e.g., `Make a decision', `Brainstorm ideas') and question-oriented (e.g., `What do we need to agree on?', `How can we think differently?') alternatives. Success criteria phrasing was preferred by pilot participants for specifying concrete results, offering a benchmark for effectiveness, and positive tone. Response options were drawn from prior meeting taxonomies (§\ref{sec:taxonomies}) and pilot feedback.

The survey also linked goal selection with outcome expectations, providing a second reflective nudge to consider goal importance \cite{thaler2008nudge,Banerjee2024NudgePlus}. Participants were asked to indicate perceived clarity of goal communication before the meeting and their \textit{expected} meeting effectiveness on a six-point scale For reasons of scope, data for these latter questions are not analysed here. Participants could also provide comments or feedback in an open-text field. Aside from text fields, all three questions were required before survey submission.

\subsubsection{Measurement: Post-meeting surveys on meeting effectiveness}\label{subsubsec:postsurveydesign}

The post-meeting surveys asked both treatment and control participants to rate meeting effectiveness on a six-point scale from \textit{`not at all'} to \textit{`extremely'} effective (right panel in \autoref{fig:surveys}). Six options were chosen to discourage default middle responses and to increase outcome spread. We also chose response options that were more positively worded to counteract the positive skew typically found in rating scales \cite{hosseinkashi_meeting_2024}, especially in collaborative contexts where participants may hesitate to rate co-workers negatively (despite ratings being private).

Participants also indicated whether the meeting had an agenda and how it was communicated (not reported here due to scope), and could optionally provide comments in an open-text field. Both closed questions were required for submission.

\subsubsection{Technical implementation} 

The app was implemented as a Microsoft Teams Bot, registered via the Bot Framework channel with a public messaging webhook URL and enabled for the Teams channel. It accessed participants' calendars using the Microsoft Graph API. The backend, deployed as a .NET 7 Kestrel web application, exposed endpoints for Teams messaging, authentication, and client UI extensions. Survey scheduling logic used the Graph API to retrieve calendar metadata, assess meeting eligibility (see criteria below), and trigger surveys at appropriate times. A SQL database recorded calendar and survey activity. Surveys were delivered as Adaptive Cards in dedicated Teams chat threads at scheduled times. In addition to survey responses, the application retrieved and stored meeting metadata used in analyses: meeting date and time, scheduled duration, organiser status, attendee count, and recurrence status.

\subsection{Sample Size and Participants}\label{subsec:sample}

Following ethics authorisation\footnote{Ethics authorization was provided by Microsoft Research’s Institutional Review Board(IORG0008066, IRB00009672).}, we recruited participants from diverse work areas within a global technology company via a company-maintained list of prior research participants and responses to internal advertisements and emails. To encourage participation, a charitable donation was made per participant to a humanitarian organisation. Participants provided consent before joining the study.

We targeted a total sample size of 400 (200 per condition) who completed the onboarding survey, installed the Meeting Survey application, and received at least one meeting survey. This was based on a power calculation using a pilot dataset (n = 41); see Appendix \ref{app:powercalcs} for details.

\begin{table}
\footnotesize
  \caption{Participant sample sizes and attrition at different study phases.}
  \label{tab:attrition}
  \begin{tabular}{llll}
    \toprule
     & Overall & Control & Treatment\\
    \midrule
     Completed consent form & 496 & - & -\\
    Completed (part of) onboarding survey & 447 & - & - \\
   Installed Meeting Survey app & 374 & 202  & 172 \\
   Completed at least 1 post-meeting survey & 361 & 198 & 163\\
    \bottomrule
  \end{tabular}
\end{table}

After attrition, the final sample included 361 participants (treatment = 163, control = 198) and 7196 meetings (see \autoref{tab:attrition} for sample size and attrition across study stages). We deviated from our preregistration by aligning inclusion criteria across groups: requiring at least one post-meeting survey (rather than one pre–post pair for treatment), adding five treatment participants (this is a less biased approach and does not affect findings; see Appendix \ref{app:exactprereg}). Participant and meeting characteristics were approximately balanced (see \autoref{tab:balance}, \autoref{tab:meetingcharacteristics}, and onboarding data in \autoref{fig:plot_prePostClarity}).

\renewcommand{\arraystretch}{1.3}
\begin{table}
\footnotesize
  \caption{Participant characteristics and balance across groups. `Other' work area pools a few participants working in `Research' and `Marketing/Promotion'. Values for the work and cognitive characteristic variables indicate average scores (see §\ref{subsubsec:obsurvey} for details).}
  \label{tab:balance}
\begin{tabular}{llcc}
\toprule
 & & Control & Treatment \\
\midrule
& \textbf{Total N} & \textbf{198} & \textbf{163} \\
\hline
Manager status & Individual contributor & 71\% & 69\%\\
& Manager & 29\% & 31\%\\
\hline
Work area & Product Development & 31\% & 38\%\\
& Sales & 23\% & 18\%\\
& Operations & 20\% & 16\%\\
& Customer Support & 11\% & 11\%\\
& IT & 9\% & 13\%\\
& Other & 5\% & 4\%\\
\hline
Seniority level & Senior & 58\% & 56\%\\
& Principal & 27\% & 25\%\\
& Early career & 14\% & 15\%\\
& Above Principal & 2\% & 4\%\\
\hline
Work characteristics & Work demands score & 2.41 & 2.40\\
& Frequency external meetings & 2.49 & 2.35\\
& Frequency leading meetings & 2.77 & 2.72\\
\hline
Cognitive characteristics & Work goal specificity score & 0.60 & 0.48\\
& (Lack of) pre-meditation & 1.68 & 1.72\\
\bottomrule
\end{tabular}
\end{table}
\renewcommand{\arraystretch}{1}

\begin{table}
\footnotesize
  \caption{Meeting characteristics.}
  \label{tab:meetingcharacteristics}
  \begin{tabular}{llll}
    \toprule
     & Overall & Control & Treatment\\
    \midrule
    Total N & 7196 & 3878 & 3318 \\
    \hline
     Recurring meetings & 35\% & 33\% & 37\%\\
    Organised by participant & 28\% & 28\% & 27\% \\
   Scheduled participant count (median) & 7 & 7  &  7\\
   Scheduled duration (median) & 30 min & 30 min & 30 min\\
    \bottomrule
  \end{tabular}
\end{table}

\subsection{Study Procedure}
\begin{figure*}[h]
\centering
\includegraphics[width=\textwidth]{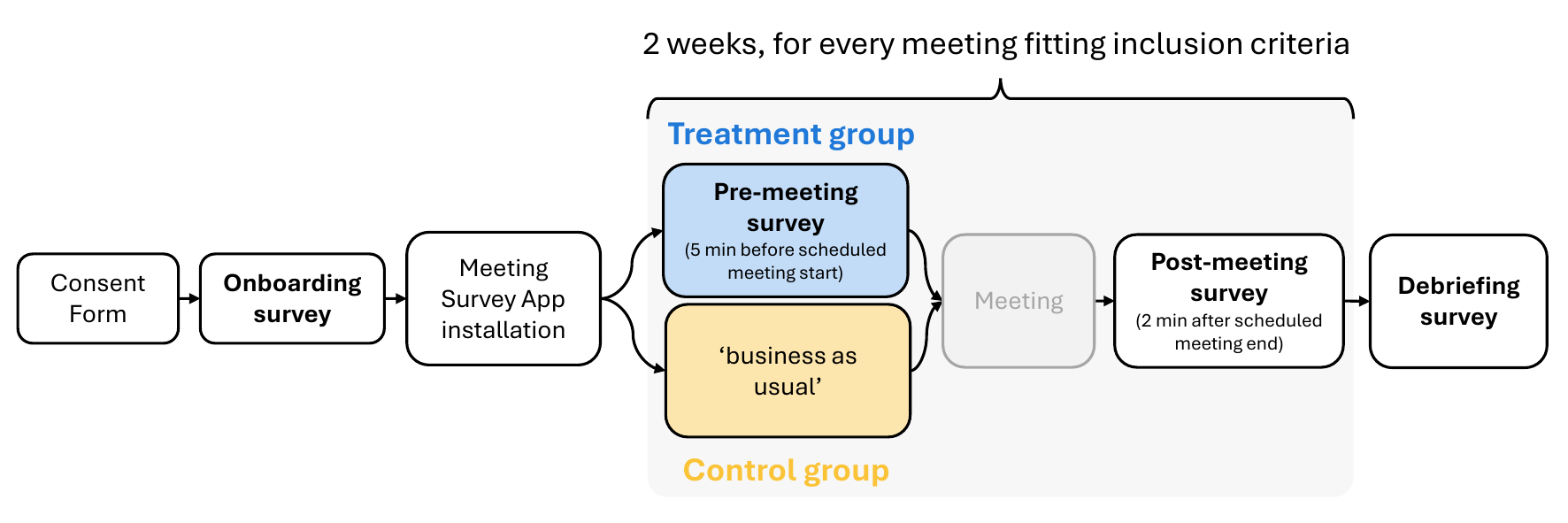}
  \caption{Participants' flow in the study. }
  \Description{Study timeline showing the flow for treatment and control groups over two weeks. The figure presents a left-to-right flow diagram of the study procedure over a two-week period, highlighting where the treatment differs from the control. Participants first complete a consent form, an onboarding survey, and install a meeting survey app. Participants are assigned to either a treatment or control condition: for each eligible meeting, the treatment group completes a brief pre-meeting survey sent five minutes before the scheduled start, while the control group proceeds with “business as usual” and no pre-meeting survey. All meetings then take place as normal. After each meeting, all participants complete a post-meeting survey sent approximately two minutes after the meeting ends. Finally, after the 2 weeks, participants complete a debriefing survey. The figure emphasizes the experimental manipulation (presence or absence of a pre-meeting survey) while keeping the rest of the workflow identical across groups.}
  \label{fig:studyflow}
\end{figure*}

After consenting, participants completed an onboarding survey (\autoref{fig:studyflow}), installed the Meeting Survey application in Teams, and continued their normal workflow. At their first eligible meeting, they were randomised to either the treatment or control group, and received the appropriate survey(s) accordingly. Surveys were delivered for eligible meetings over two weeks (10 working days), followed by a debriefing survey and application uninstall instructions. %

\subsection{Measures}
We measured a range of variables across the onboarding survey, meeting surveys, debriefing surveys, and the telemetry captured by the Meeting Survey app. We report on the subset that was included in analyses, with the rest being out of scope for the current study. Broadly, the onboarding survey (identical for both groups) asked participants about demographics, work and meeting practices, attitudes, and cognitive style. The debriefing survey asked about study experiences, ratings of pre- and post-meeting surveys, and attitudes and behaviours around goal reflection and meeting intentionality. This survey included some questions specific to the treatment group.

\subsubsection{Outcome measures}\label{subsubsec:measoutcome}
 
 Our primary outcome was self-reported \textit{meeting effectiveness}, provided per meeting via the post-meeting surveys (§\ref{subsubsec:postsurveydesign}). We opted for meeting-level measurement to improve participant recall and increase statistical power. Additionally, in the debriefing survey, we measured:
\begin{enumerate}
    \item overall meeting effectiveness and engagement (one question each, using a five-point scale from `decreased a lot' to `increased a lot')
    \item frequency of (a) feeling personally clear and (b) communicating about the meeting's goals with others, from the perspective of an organiser and attendee (one question each, with a seven-point scale from `never or almost never' to `always or almost always'; \textit{also asked in the onboarding survey for comparison})
    \item level of interest in future systems: (a) a pre-meeting prompt to think about meeting goals, (b) a post-meeting prompt to think about whether meeting goals were achieved (one question each, with a five-point scale from `not at all interested' to `extremely interested'). 
\end{enumerate}

These survey questions can be found in the Supplementary Material.

\subsubsection{Covariates}\label{subsubsec:obsurvey}

 We included a set of covariates in the analysis to increase statistical power, improve precision of our treatment estimate, and adjust for any potential imbalances between groups. We selected these as they are likely linked to our primary outcome measure of meeting effectiveness%
 . These were measured via the onboarding survey and the meeting-related telemetry from the application (for survey details see Supplementary Material). For the analysis, survey scales were recoded numerically (except for binary measures and work area); for any survey scales with multiple items, we computed an average score across items. Covariates included were: 
\begin{itemize}
    \item \textit{Work goal specificity score:} a six-item `goal specificity' sub-scale from the Conscientious Goal Setting questionnaire in \citet{bates_how_2023}, grounded in goal-setting theory \cite{locke_goal-setting_2005}. Items include 
    \textit{`I tend to have general goals rather than specific goals.'} and \textit{`I clarify my goals until they are specific'}. Participants were instructed to consider their work goals specifically. 
    \item \textit{Meeting goal clarity:} frequency of feeling personally clear on meeting goals, as an organiser and attendee (as per §\ref{subsubsec:measoutcome})
    \item \textit{(Lack of) pre-meditation:} the four-item sub-scale from the short UPPS-P trait impulsivity scale \cite{cyders_examination_2014}, focusing on the inability to consider the consequences of one's behaviour \cite{yan_cognitive_2020}. 
    \item \textit{Work demands:} four items adapted from the core `demands at work' domain in the Copenhagen Psychosocial Questionnaire, covering both quantitative demands and work pace \cite{burr_third_2019}. These were adapted to be phrased in terms of frequency over the previous 3 months.
    \item \textit{Frequency of external meetings:} a single question about the frequency of meetings involving people outside the organisation over the past three months.
    \item \textit{Frequency leading meetings:} a single question about the frequency of leading meetings in general.
    \item \textit{Manager status:} whether the participant is a manager or individual contributor.
    \item \textit{Seniority level:} participants' seniority level at work (recoded numerically).
    \item \textit{Work area:} participants' functional work area.
    \item \textit{Meeting characteristics (telemetry):} From participants' calendars via the meeting survey application, we include variables indicating whether the meeting is recurring or one-off, whether the participant organised the meeting (or an attendee), the scheduled attendee count (z-scored), and scheduled duration (z-scored).\footnote{Including `Scheduled duration' deviates from the preregistration. We include it as it captures important variance in the outcome, but this does not affect findings (see Appendix \ref{app:exactprereg}).}
\end{itemize}

\subsubsection{Descriptive measures and qualitative data}\label{subsubsec:measdescqual}
To assess intervention and study compliance, we examined and visualise the number of pre- and post- meeting surveys completed, the extent of timely completions, and the frequency and number of goals selected in the pre-meeting survey. For the mixed-methods analysis of intervention experiences, we gathered open-text feedback from participants, as well as closed question data from the debriefing survey that we visualise (see Supplementary Material). 

\subsection{Analysis}
\subsubsection{Quantitative Analysis} \label{subsubsec:quantanalysismethods}

Other than the attrition described in §\ref{subsec:sample}, for all main analyses we take an \textit{intention-to-treat} approach \cite{hollis_what_1999} that includes all data after randomisation, regardless of non-compliance. We also conducted an additional, exploratory (non-preregistered) Complier Average Causal Effect (CACE) analysis \cite{peugh_beyond_2017} that estimates intervention impact specifically for meetings where participants complied with the intervention (see Appendix \ref{app:caceresults}).

 To address missing data, we conducted multiple imputation (MI) using the \texttt{mice} package in R \cite{buuren_mice_2011} (for details on missingness and MI analyses, see Appendix \ref{app:missingness}). We also report comparable complete case analysis results for the primary outcome in Appendix \ref{app:completecaseresults}. All analyses were conducted in R.
 
 For our primary outcome, the meeting effectiveness ratings from the post-meeting surveys, we ran a mixed-effects linear model using the \texttt{lme4} package, with participant ID as a random effect, and the treatment condition and other participant-level and meeting-level variables as fixed effects (note that we deviated from our preregistration by adding scheduled meeting duration as an additional covariate). For the two outcomes measured at both onboarding and debriefing (\#2 in §\ref{subsubsec:measoutcome}), we ran a difference-in-differences model with a time by treatment interaction. For all other outcomes that were measured at the participant level, we ran standard linear models. For all analysis specifications, see Appendix \ref{app:regspecs}. 
 
 We pooled results for each imputation using Rubin's rules and conducted statistical inference using Wald t-tests on the regression coefficients. For all analyses, we report the coefficients, 95\% confidence intervals and p-values. For visualisation purposes only, we plot the raw (non-imputed) data.      

\subsubsection{Qualitative Analysis} 
To evaluate the extent and quality of participant reflection at a high level, we used GPT-4o to label each participant's open-text response as either `no change', `awareness change', `behaviour change', based on prior reflection research \cite{scott_what_2025,bentvelzen_revisiting_2022} (see Appendix \ref{app:llmdetails} for prompt details). Each label was manually reviewed by the lead author and any discrepancies corrected (see Appendix \ref{app:llmdetails}). 

Next, across all three categories, the lead author generated themes, discussed with co-authors, and manually coded all responses via reflexive thematic analysis \cite{braun_one_2021,braun_using_2006}. As authors embedded in the same organisation as participants, we situated the analysis within shared assumptions about organisational culture, meeting practices, and norms of productivity. We attended to this positionality through a reflexive analytic stance, focusing on capturing surface-level meaning, and people’s own understanding and experiences as expressed in the data; and by remaining alert to how organisational familiarity and normative commitments to reflection could shape interpretation.

\section{Findings: Quantitative Impact Evaluation}\label{sec:findings}

We first describe participants' engagement with our intervention and compliance with our field experiment. We then report on a quantitative evaluation of the reflection intervention's impact on (a) meeting effectiveness, measured per meeting, as well as (b) overall meeting effectiveness and engagement, (c) goal clarity and communication, and (d) interest in future reflection support, all measured after two weeks in the debriefing survey. In §\ref{sec:impactqual}, to contextualise our quantitative evaluation, we present mixed-methods analyses describing participants' experiences in the study, including mechanisms and barriers to reflection on goals. 

\subsection{Intervention and Study Compliance}\label{subsec:compliance}
\begin{figure}[h]
\centering
\includegraphics[width=1\columnwidth]{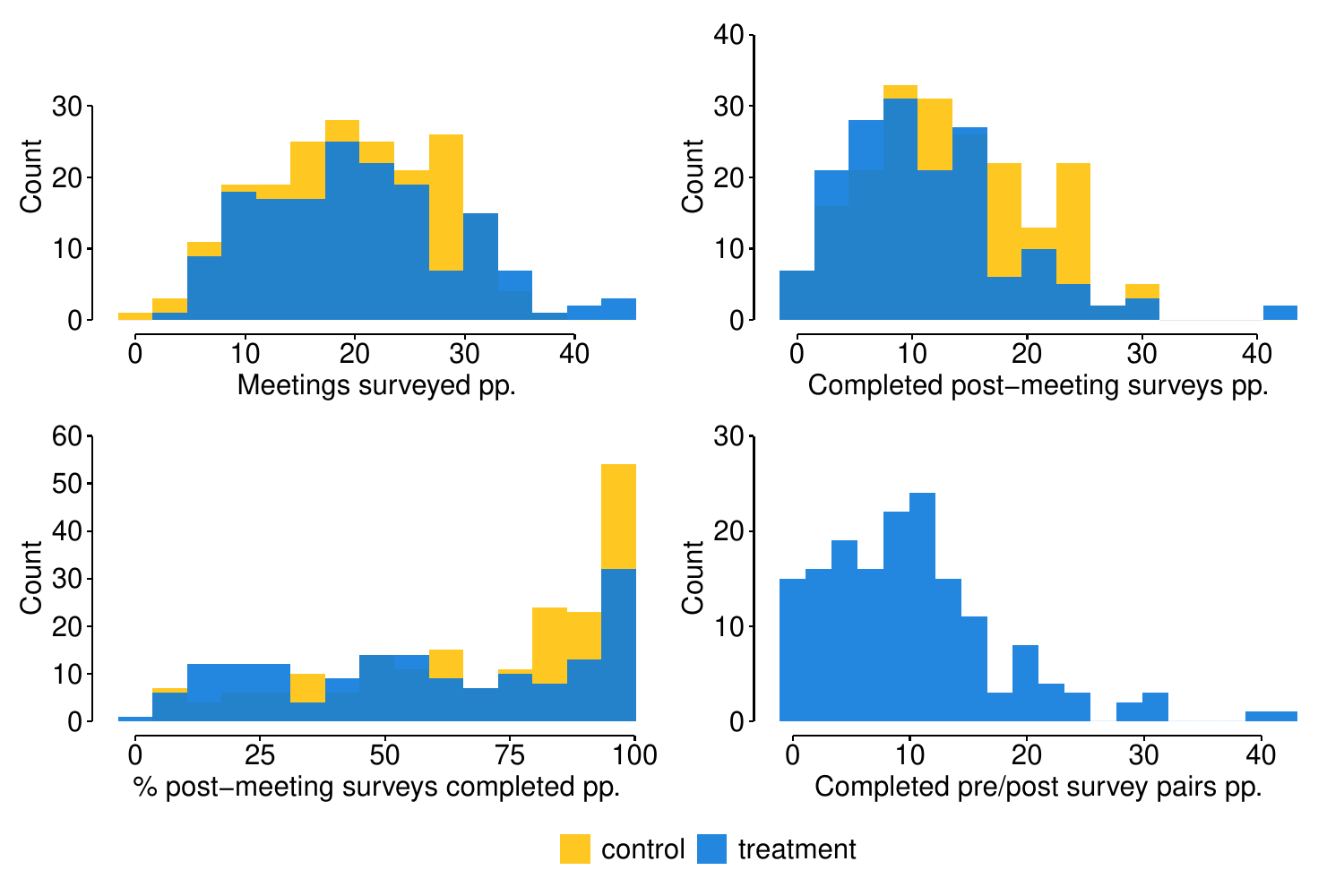}
  \caption{Distributions of intervention engagement and study compliance. \textit{Top-left:} number of meetings targeted per person by our system, by group. \textit{Top-right:} Number of completed post-meeting surveys per person, by group. \textit{Bottom-left:} Percentage of completed post-meeting surveys per person, by group. Denominators vary widely, according to the number of surveys targeted for each person. \textit{Bottom-right:} Number of completed pre- and post- meeting survey pairs per person in the treatment group.}
  \Description{Survey participation and completion across treatment and control participants. The figure shows four histograms summarizing how many meeting surveys participants completed, comparing treatment (blue) and control (yellow) groups. The top-left panel shows the number of meetings surveyed per participant, with both groups clustered around roughly 10–30 meetings. The top-right panel shows completed post-meeting surveys per participant, again with similar distributions across groups but slightly fewer completions than meetings surveyed. The bottom-left panel shows the percentage of post-meeting surveys completed per participant, with many participants—especially in the control group—completing a high proportion of surveys, including a visible concentration near 100\%. The bottom-right panel shows the number of completed pre/post survey pairs per participant (treatment only), indicating that most treatment participants contributed fewer than about 15 matched pairs, with a long tail of higher engagement.}
  \label{fig:plotSurveyCompliance}
\end{figure}

 Our data reflects a diversity in people's meeting loads and types and capacities to engage with the study, characteristic of a field experiment \cite{eden_field_2017}. The Meeting Survey application targeted an average of 20 meetings per participant over two weeks (approximately equal across groups: treatment = 3318; control = 3878; total = 7196). Meetings were unevenly distributed, ranging from 1 to over 40 per participant, reflecting varied experiences and meeting intensity (see \textit{top-left} panel in \autoref{fig:plotSurveyCompliance}).\footnote{Numbers are indicative and not precise due to occasional targeting of calendar holds and exclusion of back-to-back meetings. Overall, the numbers here are likely \textit{underestimates} of meeting frequencies.} Treatment participants completed an average of 11.2 pre/post-meeting survey pairs; control participants completed 13.4 post-meeting surveys (see \autoref{fig:plotSurveyCompliance} for distributions).
 
\begin{figure}[h]
\centering
\includegraphics[width=1\columnwidth]{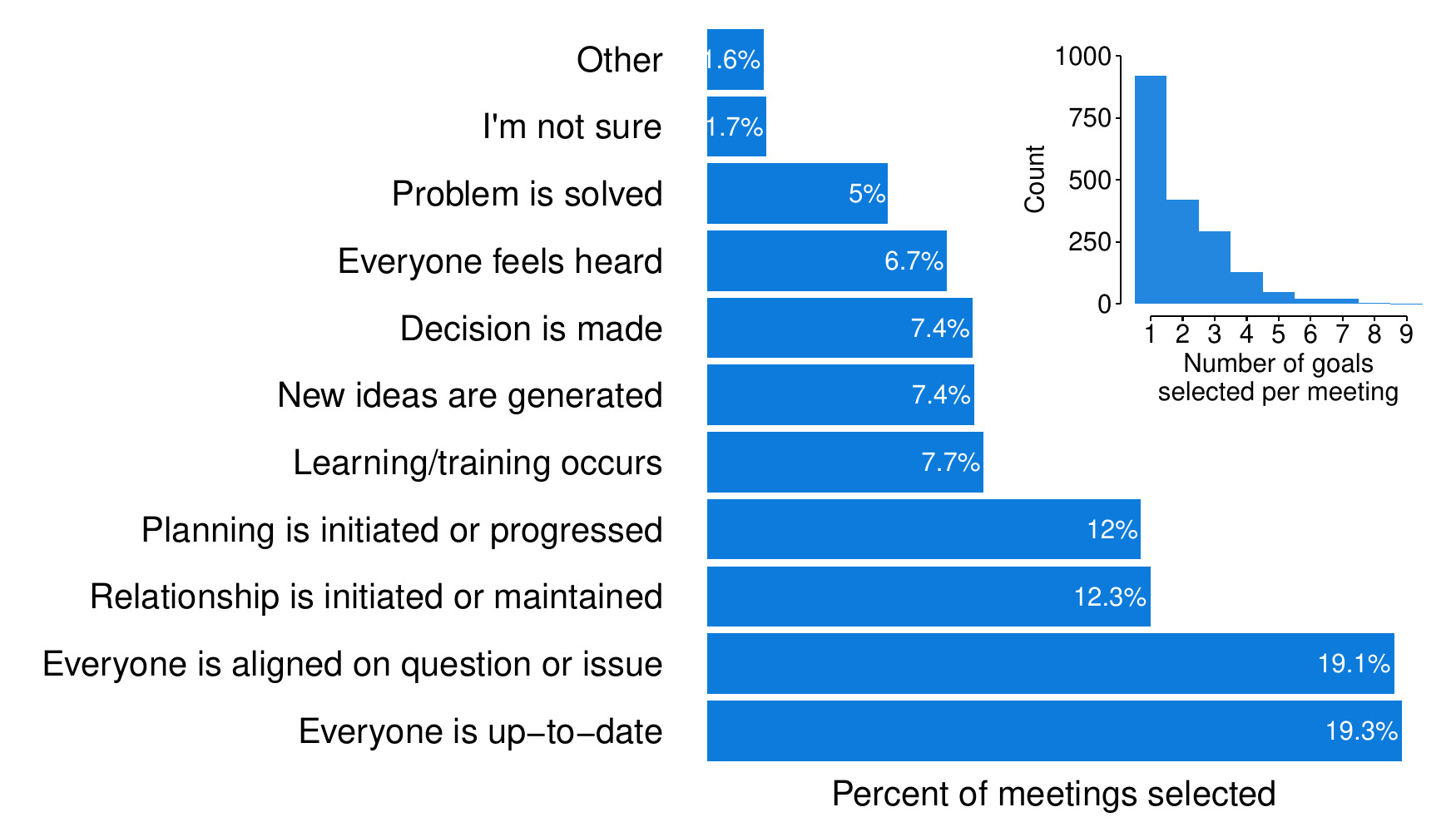}
  \caption{Type and number of goals selected across meetings in the treatment group. For the type of goals, percentages do not add up to 100\% because participants could select multiple goals for a given meeting.}
  \Description{Distribution of meeting goals selected in pre-meeting surveys. The figure summarizes which goals participants selected when asked what success looked like for their meetings, and how many goals were typically chosen. The main horizontal bar chart shows the percentage of meetings in which each goal was selected, ordered from least to most frequent. The most common goals were “Everyone is up-to-date” (19.3\%) and “Everyone is aligned on question or issue” (19.1\%), followed by relationship maintenance and planning progress (both about 12\%). Mid-range goals included learning or training, generating new ideas, and making decisions (around 7–8\%), while feeling heard, problem solving, uncertainty, and “other” goals were least frequent (under 7\%). An inset histogram on the right shows the number of goals selected per meeting, indicating that most meetings included one or two goals, with steadily fewer meetings listing three or more.}
  \label{fig:goals}
\end{figure}

In pre-meeting surveys encouraging reflection on goals, treatment participants most commonly selected \textit{`Everyone is up to date'}, and \textit{`Everyone is aligned on question or issue'} (\autoref{fig:goals}). Although multiple goals could be selected, 50\% of meetings in the treatment group had only one goal selected, which could suggest a clarity and prioritisation of goals among participants (see inset distribution in \autoref{fig:goals}). These findings indicate substantial engagement with the intervention. 
 
As expected in field experiments, especially in work settings \cite{eden_field_2017}, we also observed challenges with study compliance \cite{gerber2012field}. Participants' completion rates varied (see \textit{bottom-left} panel in \autoref{fig:plotSurveyCompliance}). Post-meeting survey completion was lower for recurring, larger, and longer meetings and also in the treatment group, with condition assignment predicting missingness in meeting effectiveness ratings (see Appendix \ref{app:missingness} and §\ref{sec:whichmeetings}). Survey response times showed 55\% of control and 60\% of treatment post-meeting surveys were completed >15 minutes after meetings, despite instructions to respond promptly (median completion time: \textit{control} = 20 minutes; \textit{treatment} = 31 minutes). For pre-meeting surveys, 55\% were completed >15 minutes after delivery, i.e., sometimes during or after the meeting (median completion time: 26 minutes). See \autoref{fig:surveycompletiontimes} in Appendix \ref{app:surveycompletiontimes} for distribution plots of completion times.

The lower completion rate and longer response time among treatment participants may have been due to the pre-meeting time being a point of greater work demand (see §\ref{subsubsec:relativetiming}). Additionally, the treatment group had to respond to twice as many surveys, which may have led to greater fatigue (see §\ref{sec:whichmeetings}).   

Despite these challenges, the dataset remains rich and relatively unique, supporting robust evaluation of the pre-meeting reflection intervention and insights into participants’ in situ experiences. We turn to this in the following sections.   

\subsection{Quantitative Evaluation of Intervention Impact Between Conditions}\label{subsec:impactquant}

\subsubsection{Meeting Effectiveness Ratings}
 We first examine the impact of the intervention on our primary outcome, meeting effectiveness ratings (at the meeting level), using mixed-effects linear regression. We do not observe a statistically significant treatment effect (p = 0.36 for the t-test on the treatment coefficient; see \autoref{fig:meetingEff} for all coefficients, 95\% confidence intervals, and p-values).\footnote{See Appendix \ref{app:meetingEffExtras} for an interpretation of other correlates of self-reported meeting effectiveness in \autoref{fig:meetingEff}}. A complete case analysis and an analysis without preregistration deviations showed the same results (Appendix \ref{app:completecaseresults} and \ref{app:exactprereg}). Given the high proportion of pre-meeting surveys completed late or not at all (§\ref{subsec:compliance}), alongside the \textit{intention-to-treat} analysis above, we also conducted an exploratory (not preregistered) \textit{Complier Average Causal Effect (CACE)} analysis \cite{peugh_beyond_2017} that estimates intervention impact specifically for meetings where participants complied with the intervention. This analysis also did not show a statistically significant treatment effect (p = 0.6; see Appendix \ref{app:caceresults}). Thus, we do not find evidence that pre-meeting goal reflection improved self-reported meeting effectiveness measured at the meeting level.    

\begin{figure*}[h]
\centering
  \includegraphics[width=\textwidth]{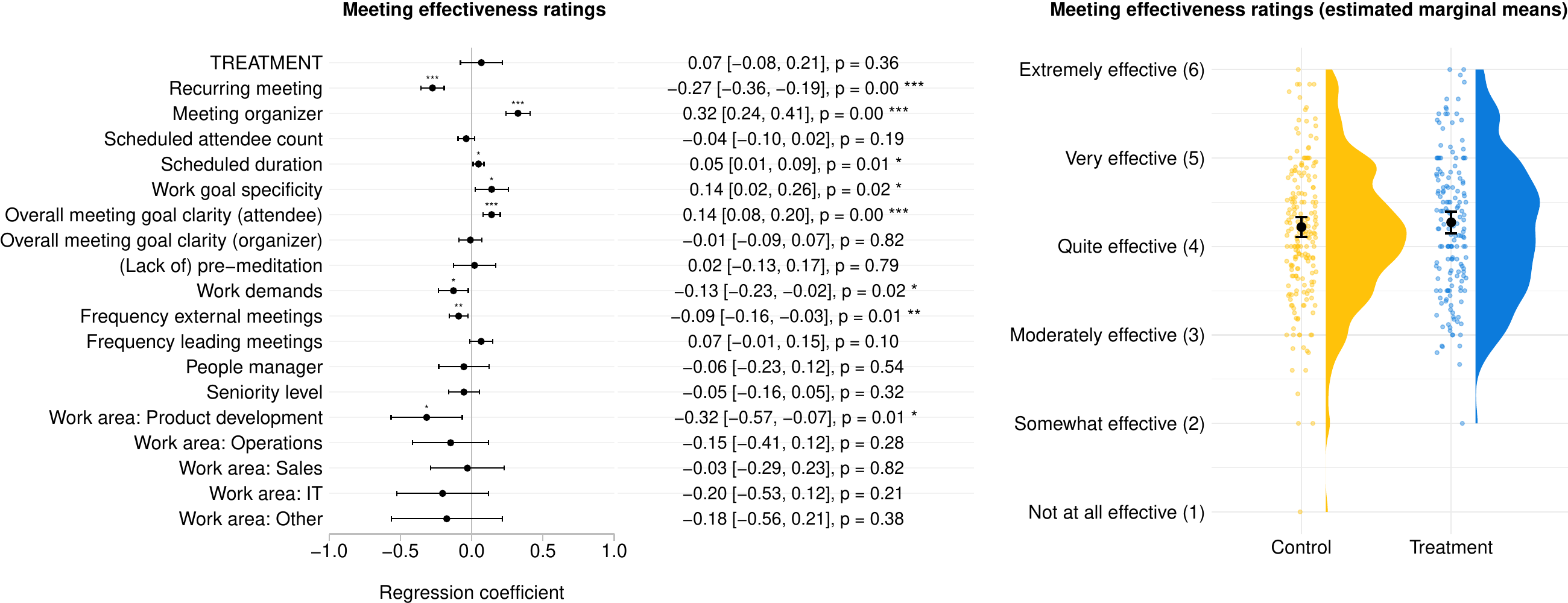}
  \caption{Analysis results for the meeting effectiveness ratings. Left panel displays regression coefficients (black markers) for the condition assignment (`TREATMENT') and the covariates, along with error bars indicating 95\% confidence intervals (CIs). Positive coefficients indicate a positive association with meeting effectiveness ratings, and vice versa. The numerical values to the right show the coefficients, along with 95\% CIs and p-values. Statistical significance indicators: ** p<0.001, ** p<0.01, * p<0.05, + p<0.1. The right panel visualises participants' average meeting ratings (coloured markers) for each condition, as estimated marginal means derived from a linear mixed model on the complete-case data (for visualisation only). To the right are density distributions of the participant data. Black markers indicate group averages and 95\% CIs.}
 \Description{Description.}
 \label{fig:meetingEff}
\end{figure*}

\subsubsection{Overall Meeting Effectiveness and Engagement}

\begin{figure}[h]
\centering
  \includegraphics[width=1\columnwidth]{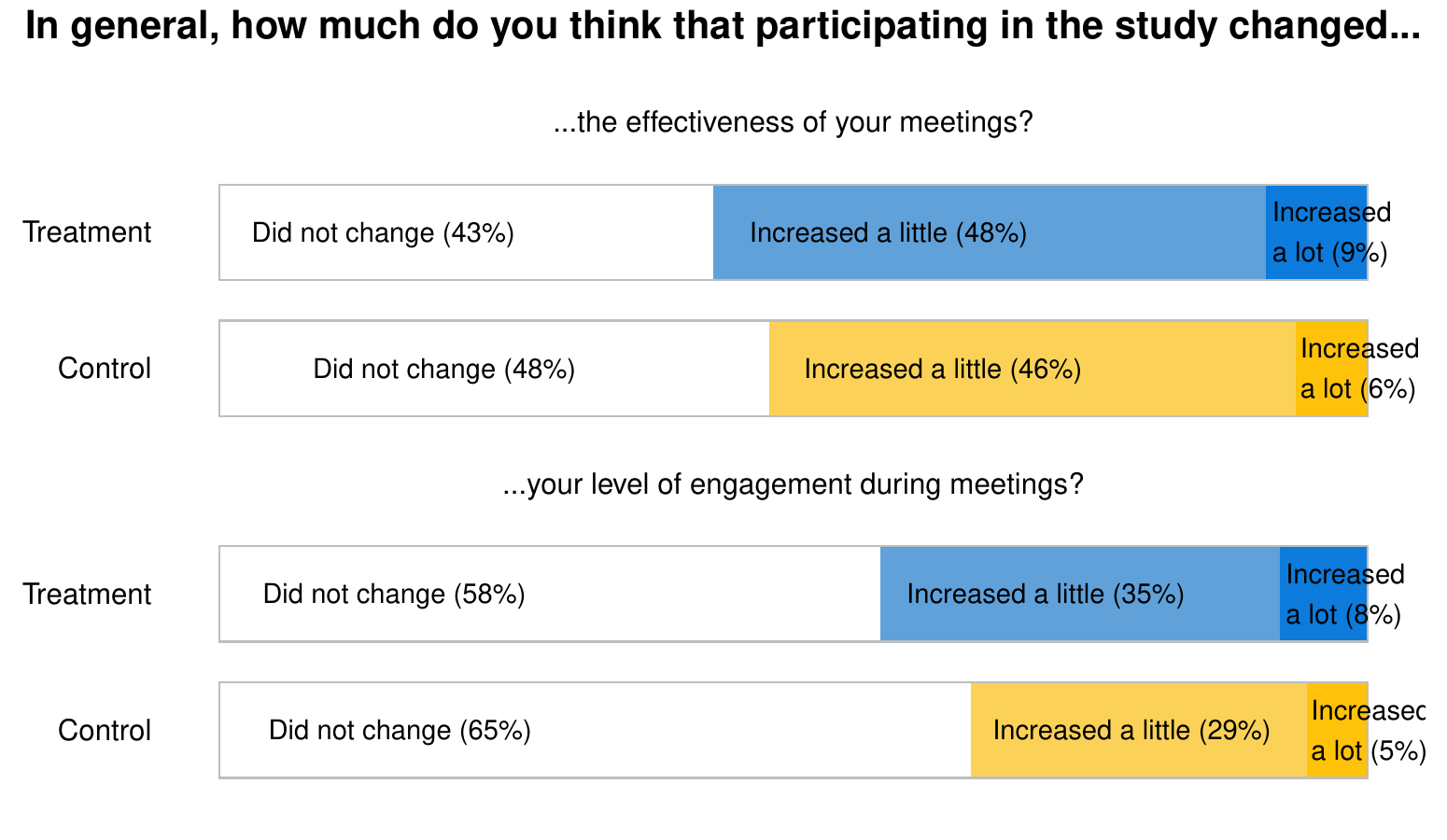}
  \caption{Overall meeting effectiveness (top) and engagement (bottom), measured in the debriefing survey. Plots show the raw data as proportions of responses in each category. Note that response options included `Decreased a little' and `Decreased a lot' but these did not receive any responses and are not shown.}
 \Description{Self-reported impact of study participation on overall meeting effectiveness and engagement across two weeks. The figure shows stacked horizontal bars comparing treatment and control participants’ perceptions of how participating in the study affected their meetings. The top pair of bars reports perceived changes in meeting effectiveness: in both groups, about half of participants reported no change, while the remainder mostly reported that effectiveness “increased a little,” with a small minority reporting it “increased a lot” (9\% in treatment, 6\% in control). The bottom pair reports perceived changes in engagement during meetings, showing a similar pattern but with more participants in both groups reporting no change, especially in the control group. Across both outcomes, treatment participants are slightly more likely than controls to report increases, but the dominant response in all cases is little or no perceived change.}
 \label{fig:genMeetingEffMeetingEng}
\end{figure}

Although the meeting-level analysis showed no significant intervention effect, aggregate impacts may still occur. The intervention did not target all meetings (e.g., we excluded back-to-back meetings), yet its longitudinal design could shape behaviour beyond those measured. Participants might also fail to notice changes in individual meetings but still experience a cumulative shift over time. As such, we measured and analysed \textit{overall} meeting effectiveness and engagement, reported after two weeks in the debriefing survey. Compared to 52\% in the control group, 57\% of those in the treatment group reported that study participation increased the effectiveness of their meetings `a little' or a `a lot'; however, this difference is not statistically significant (p = 0.13 for the treatment coefficient in a linear regression; \autoref{fig:genMeetingEffMeetingEng} left panel). For engagement, the analogous figures were 34\% in the control group compared to 43\% of participants in the treatment group, a larger difference but which was also not statistically significant (p = 0.05 for the treatment coefficient in a linear regression; \autoref{fig:genMeetingEffMeetingEng} right panel). See Appendix \ref{app:overallmeetingeffeng} for full results. Thus, we do not find evidence that pre-meeting goal reflection improved overall self-reported meeting effectiveness and engagement, measured after study participation.

\subsubsection{Goal Clarity and Communication} \label{subsubsec:goalclarityfindings}
Meeting effectiveness and engagement are \textit{collaborative} outcomes ultimately depending on all attendees, including those outside our study. Since the intervention targeted \textit{individual} volunteers, its effects may be more likely to appear in their own self-reported \textit{goal clarity} and \textit{communication}, which indirectly shape broader meeting effectiveness and engagement. We analysed participants' self-reports of how often they felt (a) personally clear on the meeting goals, and (b) communicated about the meeting goals with others, as meeting organisers and attendees. As these outcomes were measured both at onboarding and debriefing, we analysed them using difference-in-difference models, comparing before and after the study, and treatment and control groups. For all four outcomes, although there was no significant interaction between treatment and time (all p-values > 0.1), there was a \textit{significant main effect of time} (all p-values < 0.05), with participants in both groups reporting improved goal clarity and communication \textit{after the study} (see Appendix \ref{app:goalcomm} for details). \autoref{fig:plot_prePostClarity} illustrates that self-reported frequencies increase at the end of the study (more so for organiser than attendee perspectives), but that this improvement occurs for \textit{both} treatment and control groups. Thus, these findings suggest that both treatment and control group experienced comparable changes in meeting goal clarity and communication. The mixed-methods evaluation below explores how the post-meeting surveys may have contributed to the observed changes in the control group (§\ref{subsec:comparablechanges}).

\begin{figure*}[h]
\centering
  \includegraphics[width=\textwidth]{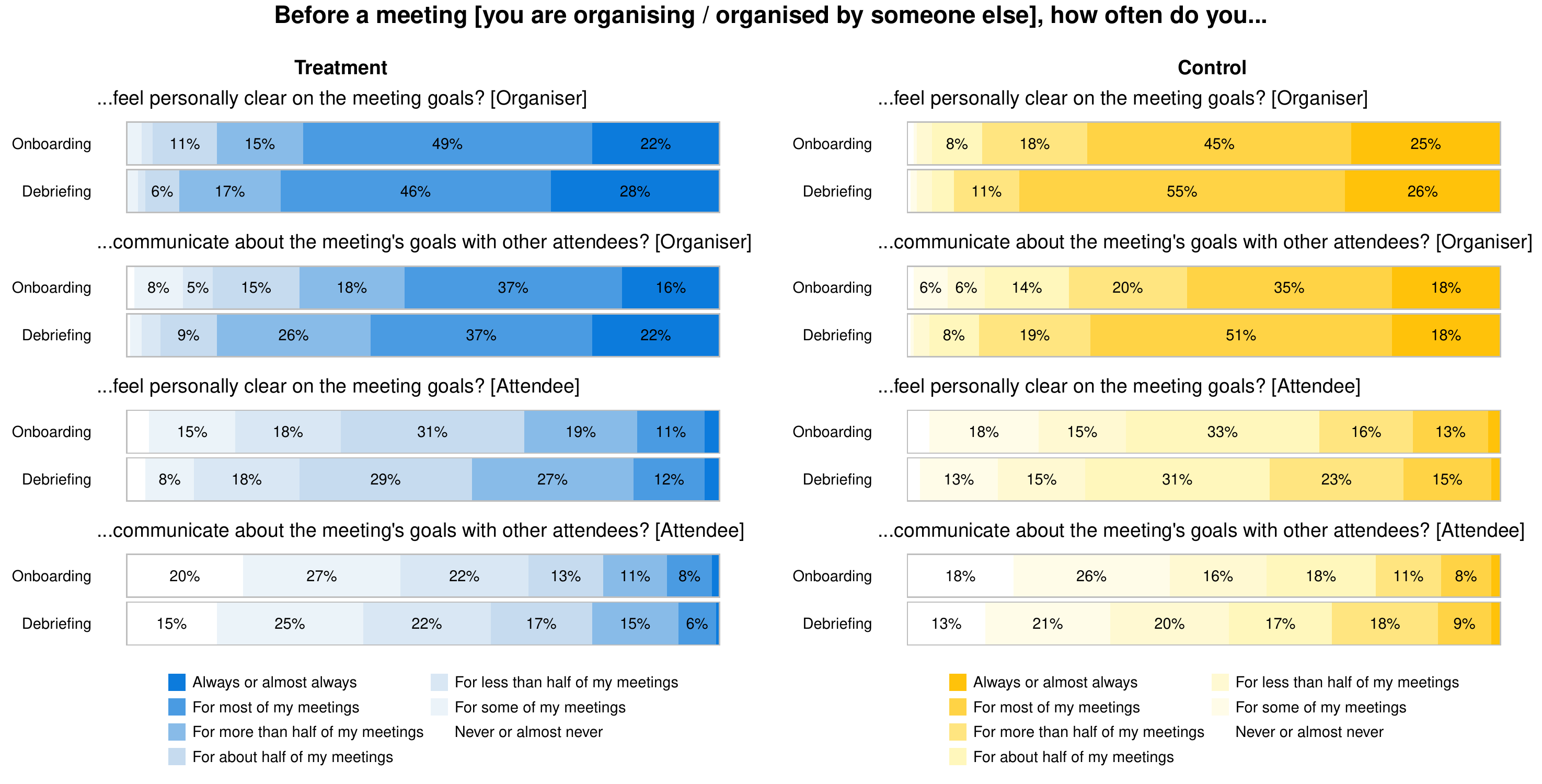}
  \caption{Frequency of meeting goal clarity and communication from the perspective of an organiser and attendee, measured in the onboarding and debriefing surveys, for treatment and control groups. Raw data visualised as proportions of responses in each category. }
 \Description{Description.}
 \label{fig:plot_prePostClarity}
\end{figure*}

\subsubsection{Interest in Future Reflection Support}
\begin{figure}[t]
\centering
\includegraphics[width=1\columnwidth]{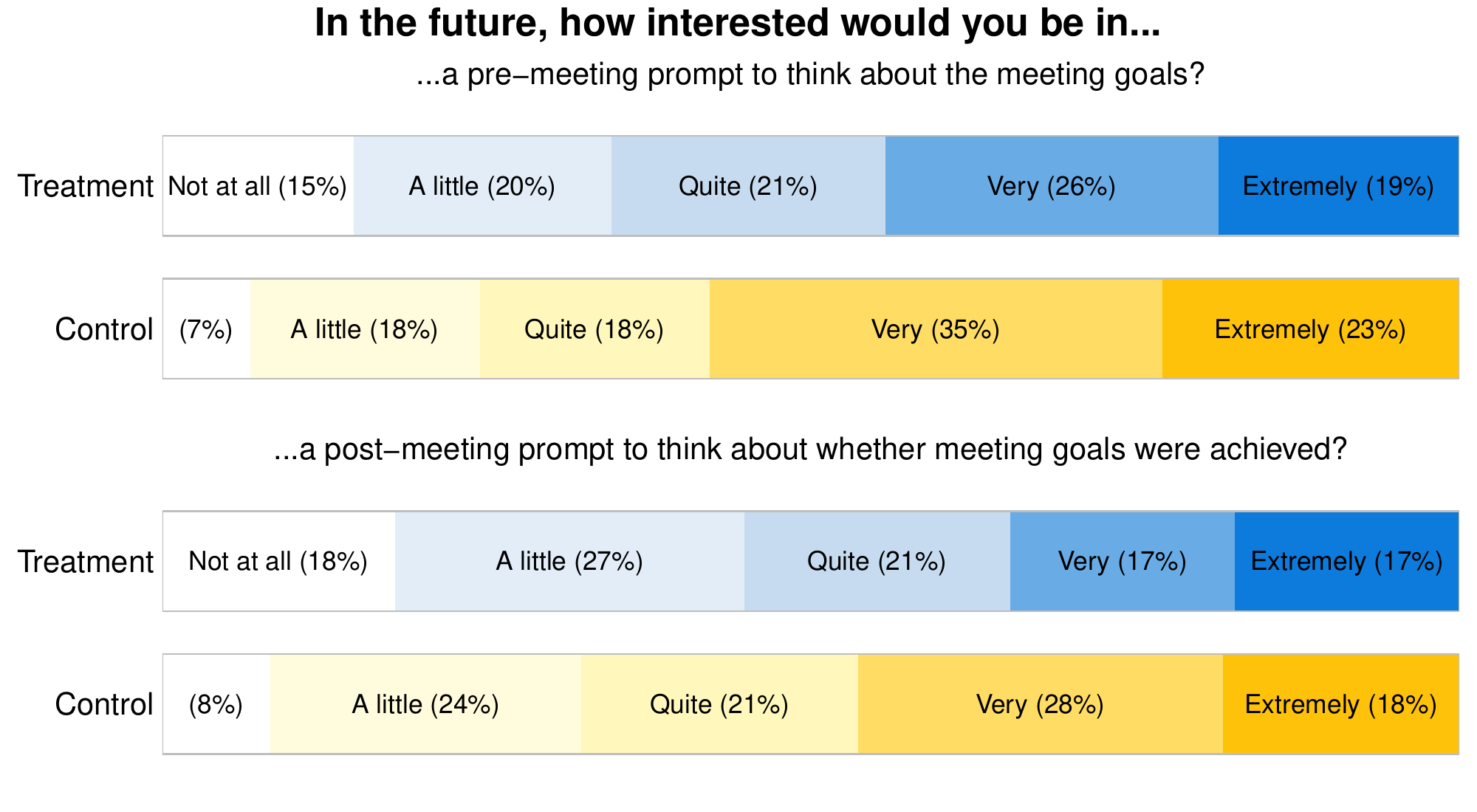}
  \caption{Interest in future reflection support: pre- and post- meeting prompts to support goal reflection. Raw data visualised as proportions of responses in each category. Participants in the treatment group were on average less interested than those in the control group.}
  \Description{Interest in future pre- and post-meeting reflection prompts. The figure shows stacked bars summarizing how interested treatment and control participants would be in using meeting reflection prompts in the future. The top pair of bars reports interest in a pre-meeting prompt to think about meeting goals: in both groups, most participants express moderate to high interest, with treatment responses spread across “quite,” “very,” and “extremely,” and control participants showing a larger share selecting “very” or “extremely interested.” The bottom pair reports interest in a post-meeting prompt to reflect on whether goals were achieved, again showing generally positive interest in both groups, though with slightly more interest in the control group compared to the treatment group. Overall, the figure indicates broad openness to both pre- and post-meeting reflection prompts, with interest among the control group slightly higher than the treatment group.}
  \label{fig:interestPrePost}
\end{figure}

Finally, after their two-week experience, we sought to understand participants' interest in further reflection support, as another indicator of impact. We asked both groups in the debriefing survey about their future interest in a \textit{pre-meeting prompt} to help them think about meeting goals. Over two-thirds were `quite', `very', or `extremely' interested in this (\autoref{fig:interestPrePost}). However, the treatment group was \textit{less} interested than the control group (p = 0.01 for the treatment coefficient in a linear regression; see Appendix \ref{app:interestPrePrompt} for details). We also asked about interest in a \textit{post-meeting prompt} to help think about whether meeting goals were achieved. Again, despite broad interest overall, the treatment group was less interested than the control group (p = 0.03 for the treatment coefficient; Appendix \ref{app:interestPrePrompt}). These unexpected differences suggested that survey fatigue may have impacted the treatment group experience. In the mixed-methods evaluation below, we turn to this and other aspects of the study experience to understand the lack of hypothesised intervention impact.

\section{Findings: Mixed-Methods Evaluation of Experiences, Mechanisms, and Barriers}\label{sec:impactqual}

We conducted a mixed-methods evaluation to understand participants’ experiences during the two-week study and to identify mechanisms and barriers to change. Findings draw on qualitative open-text analysis and descriptive closed-question survey data about participants’ perceptions. We frame our findings around themes explaining the null treatment effect in the quantitative evaluation. We interpret the null effect as potentially arising from the combination of two factors that organise the following sections:

\begin{enumerate}
    \item The pre-meeting survey intervention had \textit{less impact than expected}. This potentially arose due to intervention design, behavioural, and socio-cultural challenges making pre-meeting reflection less frequent or less meaningful for participants, or precluding it from having downstream impact on meeting effectiveness (§\ref{subsec:lackchanges}).
    
    \item The post-meeting surveys had \textit{more impact than expected}, inadvertently functioning as a reflection intervention sufficient to trigger comparable changes in the control group (evidenced by quantitative findings in §\ref{subsubsec:goalclarityfindings} and qualitative findings below). This potentially arose due to the regularity of the surveys, as well as their focus on goals, which made reflection meaningful and actionable (§\ref{subsec:comparablechanges}).  
\end{enumerate}

To explore these two factors, we analysed participants' open-text responses, labeling them as either `\textit{no change}', `\textit{awareness change}', or `\textit{behaviour change}'. %
The proportions of these labels were comparable between the treatment and control groups ($\chi^2$ = 1.97, p = 0.37; see \autoref{tab:labelprops}). Notably, changes in awareness and behaviour were reported by a majority of participants in both groups (57\% of treatment and 60\% of the control; \autoref{tab:labelprops}), echoing the quantitative evaluation findings above (§\ref{subsubsec:goalclarityfindings}). We then thematically analysed responses in each category and complemented this with descriptive closed-question survey data. We report these findings below.

\begin{table}
  \caption{Proportion of participant open-text responses per category. Some participants did not provide open-text feedback, reducing the totals here compared to the overall sample.}
  \label{tab:labelprops}
  \begin{tabular}{lll}
    \toprule
      & Control & Treatment\\
    \midrule
    Total N & 187 & 155 \\
    \hline 
     No change  & 40\% & 43\%\\
    Awareness change & 41\% & 34\% \\
   Behaviour change  & 19\%  &  23\%\\
    \bottomrule
  \end{tabular}
\end{table}

\subsection{Potential Reasons for Pre-Meeting Surveys' Decreased Impact}\label{subsec:lackchanges}

We identified five themes suggesting why the pre-meeting intervention may have had limited impact. These include the targeting of meetings for reflection; the timing of reflection prompts relative to the meeting; the interface and modality for reflection, challenges with the inherent interdependence of collaboration; and broader work culture barriers to reflection. 

\subsubsection{Targeting the `Right' Meetings.} 
\label{sec:whichmeetings}
\begin{figure*}[t]
\centering
\includegraphics[width=\textwidth]{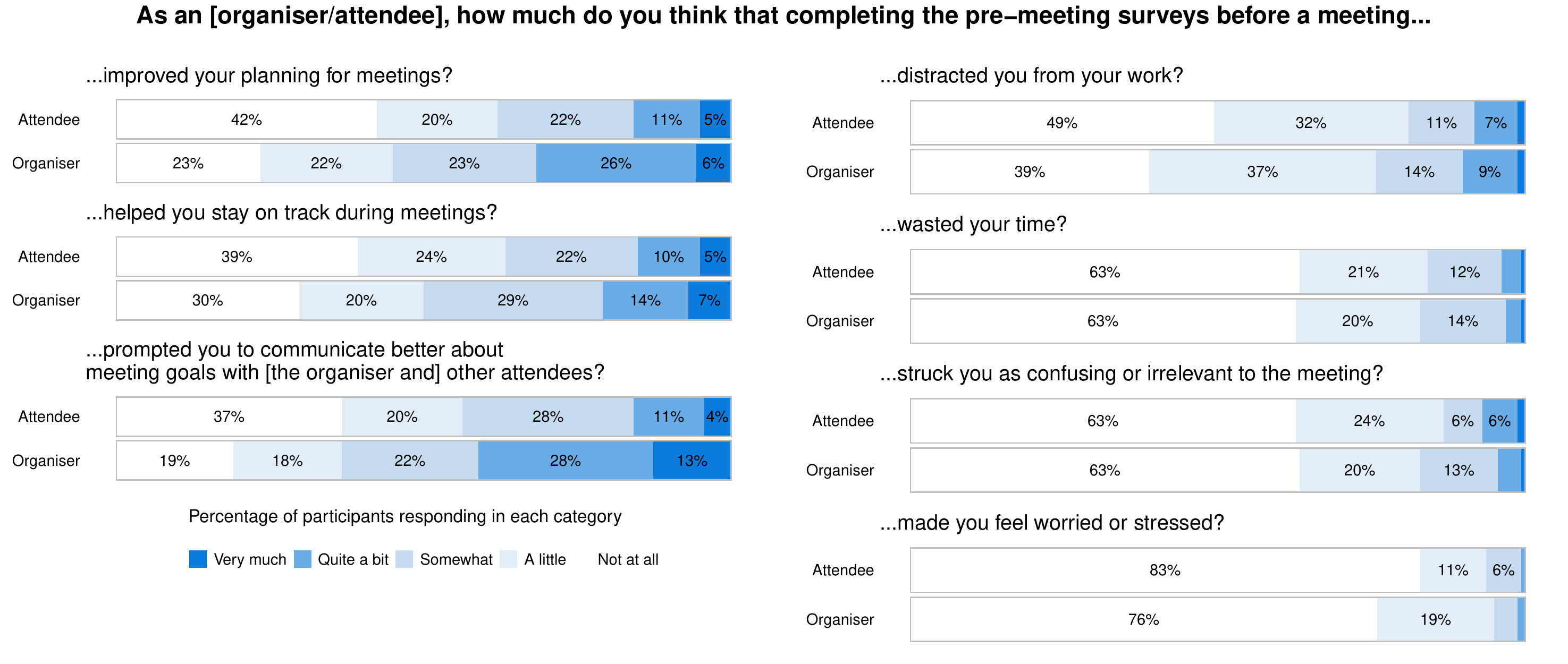}
  \caption{Perceived positive and negative impacts of the pre-meeting surveys. Participants were asked to answer the same set of questions from the perspective of meeting organisers and attendees. Raw data is visualised as proportions
of responses in each category.}
  \Description{Perceived benefits and downsides of completing pre-meeting surveys. The figure shows stacked horizontal bars summarizing how participants in the treatment group perceived the impact of completing pre-meeting surveys as attendees and organizers. The left column focuses on potential benefits: improving meeting planning, helping participants stay on track during meetings, and prompting better communication about meeting goals. Across these outcomes, organizers report stronger positive effects than attendees, with sizable shares indicating the surveys helped “quite a bit” or “very much,” especially for planning and goal communication. The right column shows potential downsides: distraction from work, wasting time, being confusing or irrelevant, and causing worry or stress. For both roles, most respondents report “not at all” or only “a little” negative impact, with very small proportions reporting substantial downsides. Overall, the figure indicates that pre-meeting surveys are perceived as more helpful than harmful, particularly for organizers, with minimal reported costs.}
  \label{fig:preeff_posneg}
\end{figure*}

A key challenge concerned \textit{which} meetings to target for reflection. Participants described being fatigued by the amount of surveys they received \pid{(T436, T308, T346, T282, T204, T127, T446, T504)}. Between about 10-25\% of participants found the pre-meeting surveys to be at least `somewhat' confusing, irrelevant, distracting, or wasting their time (\textit{right panel} in \autoref{fig:preeff_posneg}).  

\begin{smallquote}
     \iquote{``Having a survey for EVERY meeting when I have a lot of back-to-back meetings can feel very tedious after a while.''} \pid{(T308)}
\end{smallquote}

Certain meetings were not considered important for reflection, such as townhalls \pid{(T204, T335)}, meetings viewed as having well-understood or implicit goals, such as ad hoc meetings \pid{(T144)}, meetings with external partners \pid{(T424)}, or catch-ups \pid{(C21)}. Those in customer support roles viewed all their meetings as having a consistent set of goals: \iquote{``identify customer's issue, troubleshoot, deliver solution''} \pid{(T202)}. \autoref{fig:barriers} shows that meetings having ``too many people'' or ``too many people with different roles'' is a barrier to goal reflection for around a quarter of participants, possibly suggesting the difficulty of expressing goals to diverse groups. Conversely, almost a tenth of participants perceived certain meetings as having too \textit{few} people for goals to be relevant.  

Some participants did not want individual prompts for multiple instances of a recurring series \pid{(T38, T308, T342, T282, T204)}. Yet, recurring series were also characterised as sometimes becoming \iquote{``divorced from their original goals''} \pid{(C255)}, and  would therefore benefit from reflection, at least by organisers \pid{(T492, C255, C42, C61)}. Indeed, recurring meetings were rated as significantly less effective in our data (see \autoref{fig:meetingEff} and Appendix \ref{app:meetingEffExtras}).

Participants suggested smarter targeting of meetings for reflection, for example, based on whether meetings are information-gathering or decision-making meetings \pid{(C89)}, their `importance' \pid{(T113)}, or the person's role \pid{(T326)}. Better targeting also concerns the relative \textit{timing} of prompts within workflows, as discussed below.

\subsubsection{Relative Timing of Reflection Prompts.} 
\label{subsubsec:relativetiming}

The timing of reflection prompts relative to the meeting was also problematic. Pre-meeting surveys were reported as being sent too close to the meeting, leaving insufficient time to respond, especially with back-to-back meetings \pid{(T308, T326, T335, T236, T220, T117, T160)} (see also §\ref{sec:culture}). This partly explains the high proportion of incomplete or late surveys reported in §\ref{subsec:compliance}. Participants also reported not having sufficient time to act, including clarifying and sharing meeting goals \pid{(T56)}, deciding whether to attend \pid{(T326)}, or contacting the organiser for more information \pid{(T236)}.%

\begin{smallquote}
     \iquote{``I prep for meetings and re-gather my thoughts, questions, and materials hours before the meeting starts. When the survey is sent 5 minutes before the meeting starts, we are already way past time when I would change anything about my approach.''} \pid{(T236)}
\end{smallquote}

Participants instead wanted prompts at the time of scheduling \pid{(T548, T540, T152, T171, T172)}, or to reflect in `batches' at timely moments \pid{(T438, T504, T93, T481, C341, C247)}, such as on \iquote{``all meetings for the day at the beginning [...] and reviewing the meetings [...] at the end of the day''} \pid{(T438)}. \textit{When} to reflect intertwines with \textit{where} and \textit{how}.

\subsubsection{Interfaces and Modalities for Reflection.} \label{sec:wherehow}

 The place and modality of our intervention was another challenge. Some participants noted that surveys in chat were easy to miss amid multiple other notifications \pid{(T73, T117, T272, T486)}. Instead, whereas some participants wanted surveys to be embedded into the videoconferencing interface \pid{(T73, T248)}, many participants highlighted the meeting scheduling interface as a key site for goal reflection \pid{(T152, T171, T172, T548, T122, T540, C296, C44, C147)}. There, some participants wanted active support with clarifying and communicating goals \pid{(T122, T172, T174, C296)}, such as via templates, checklists, or AI that can surface context and enable deeper reflection \pid{(T372, T548, T194, C68, C147)}. This could also address the difficulties some participants had with answering our closed-format surveys, related to the amount or indistinctiveness of the response options \pid{(T489, C51, C253)}. These considerations intersect with participants' roles as meeting organisers or attendees, which we turn to next.   

\begin{smallquote}
    \iquote{``...when someone creates an Outlook Meeting [...] the organiser has an 'Optional' choice to walk-through a guided process of defining the Agenda, Goals, Purposes, etc. before sending the calendar invite? Then this [...] would format an Agenda into the email body to ensure everyone knows the goals and outcomes for the meeting?''} \pid{(T122)}
\end{smallquote}

\subsubsection{The Interdependence of Collaboration} \label{sec:who}

The inherent interdependence of collaboration emerged as a key barrier in two ways. First, participants distinguished between meetings they \textit{organised}, which they saw as benefiting from reflection, and meetings to which they were \textit{invited}, which may benefit less \pid{(T88, T492, T558, T283, T242, T313, T316, T360)}---an important distinction given that only about a third of meetings in our sample were organised by the participants (\autoref{tab:meetingcharacteristics}). Indeed, participants rated the pre-meeting surveys as substantially less helpful from the perspective of attendees (\textit{left panel} in \autoref{fig:preeff_posneg}). For some attendees, this boiled down to not being \iquote{``in a position of power''} \pid{(T88)} to change meetings or ask organisers for more information \pid{(T558, T283, T98, T223)}. This organiser-attendee asymmetry mirrors that seen in the quantitative findings (\autoref{fig:plot_prePostClarity}).

\begin{smallquote}
     \iquote{``%
     I think my daily scrum (4 days a week) is a drain on my time. I've discussed this with the organiser who thinks that frequency is necessary, so I can't change much.''} \pid{(T492)}
\end{smallquote}

Second, broader than the organiser-attendee distinction, participants noted that meeting effectiveness depends on co-operation before meetings (e.g., to prepare) and during meetings (e.g., to stay on track), which is not always guaranteed \pid{(T155, T357, T460, T28)}. This points to a broader work culture challenge, discussed below.  

\begin{smallquote}
    \iquote{``I can't say [reflection] changed the actual outcome of the meetings, though. If all parties aren't prepared, meetings are still ineffective.''} \pid{(T460)}
\end{smallquote}

\subsubsection{Work Culture Barriers}\label{sec:culture}

\begin{figure*}[h]
\centering
\includegraphics[width=\textwidth]{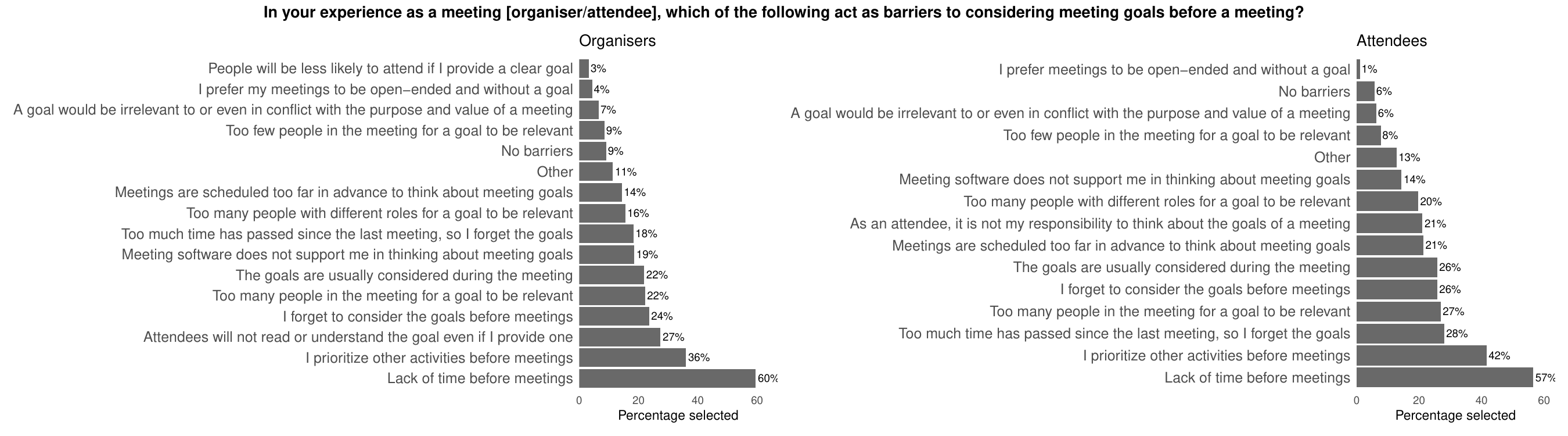}
  \caption{Perceived barriers to considering meeting goals before a meeting, for all participants. Participants were asked to answer the same set of questions from the perspective of meeting organisers and attendees. Percentages do not add up to 100\% because participants could select multiple options.}
  \Description{Barriers to considering meeting goals before meetings, by role. The figure presents two horizontal bar charts comparing organisers’ and attendees’ reported barriers to thinking about meeting goals before a meeting. Bars show the percentage selecting each barrier, ordered from least to most common. For both roles, the most frequently reported barrier is lack of time before meetings (60\% of organisers, 57\% of attendees), followed by prioritising other activities before meetings. Many also report forgetting goals due to time gaps between meetings, having too many people or roles involved for a single goal to feel relevant, and assuming goals are addressed during the meeting itself. Organisers more often cite concerns about attendees not reading or understanding stated goals, while attendees more often say it is not their responsibility or that meeting software does not support goal-setting. Very few respondents in either group report having no barriers or preferring meetings without goals.}
  \label{fig:barriers}
\end{figure*}

Participants' organisational culture, defined here by meeting practices and work intensity, served as another barrier. For some, this boiled down to a \iquote{``lack of culture in pre-read''} \pid{(T32)}, where sharing and reading preparatory materials for meetings isn't common practice. 21\% of all participants felt that, as an attendee, it was not their ``responsibility to think about the goals of a meeting'' (\textit{right panel} in \autoref{fig:barriers}). 

\begin{smallquote}
     \iquote{``People don't read agendas before attending. No pre-work is ever done, the meeting is when the content gets discussed.''} \pid{(C532)}
\end{smallquote}

Accordingly, organisers assumed that even if they shared meeting goals and other pre-reads, attendees wouldn't read them anyway \pid{(T460, T155, C367, C532)}. From the organiser perspective, 27\% of participants felt that a barrier to considering goals was that ``attendees will not read or understand the goal'' even if provided, the third most commonly reported barrier (\textit{left panel} in \autoref{fig:barriers}). Conversely, some attendees reported reaching out to organisers about goals only to get \iquote{``undetailed and unsatisfactory''} responses \pid{(T236)}.

This culture around meetings was intertwined with participants' \iquote{``fast pace of activity''} \pid{(C284)}, which left them insufficient time to prepare for meetings \pid{(T130, C532)}. %
Both organisers and attendees reported lack of time as the \textit{most} common barrier to reflecting on meeting goals, followed by a prioritisation of other tasks before meetings (\autoref{fig:barriers}). Participants therefore resisted change because \iquote{``all the organisation should be aligned''} \pid{(T28)}. %

\subsection{Potential Reasons for Post-Meeting Surveys' Increased Impact}\label{subsec:comparablechanges}

We identified three intertwined factors that explain why the control group may have experienced impact comparable to the treatment group. First, like the pre-meeting surveys, the post-meeting surveys served as regular reminders to reflect on meetings, which control participants could benefit from. More than mere reminders, however, both surveys' focus on the related concepts of goals and effectiveness served as a reflective lens sufficient to stimulate a range and depth of reflection that was also observed in the control group. Finally, the focus on goals also appeared to function as a thread connecting across workflows, sufficient to trigger actions that even control group participants reported taking. We expand on these below.          

\subsubsection{A Habitual Reminder to Reflect}\label{subsubsec:comparable-finding-habitual}
\begin{figure*}[t]
\centering
\includegraphics[width=\textwidth]{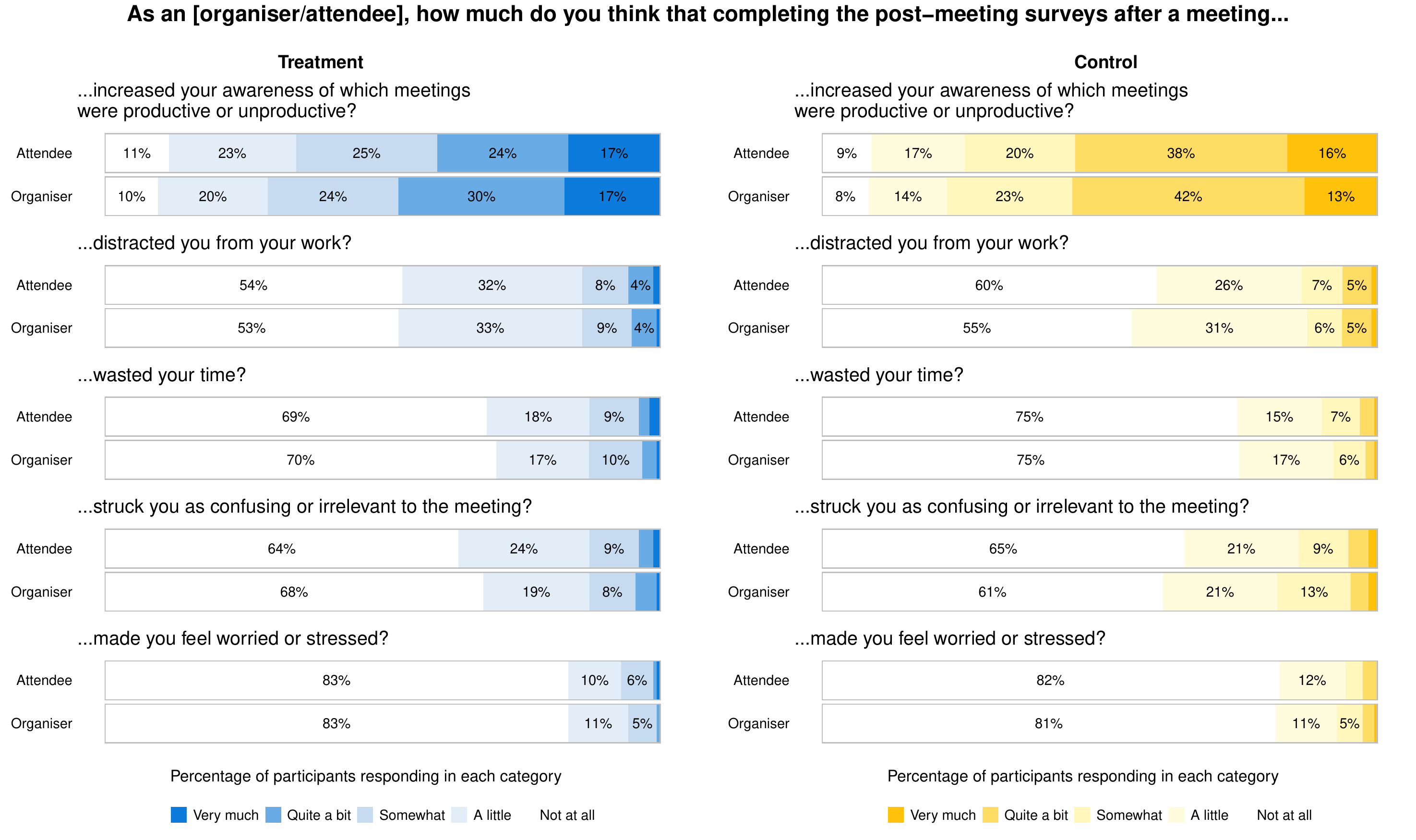}
  \caption{Perceived positive and negative impacts of the post-meeting surveys for treatment (left) and control (right) group participants. Participants were asked to answer the same set of questions from the perspective of meeting organisers and attendees. Raw data is visualised as proportions of responses in each category. There were no significant differences between groups.}
  \Description{Perceived effects of completing post-meeting surveys, by role and condition. The figure shows stacked horizontal bars comparing treatment and control participants’ perceptions of post-meeting surveys, separately for attendees and organizers. The top row shows perceived benefits: in both groups, many participants report that post-meeting surveys increased their awareness of which meetings were productive or unproductive, with sizable shares selecting “quite a bit” or “very much,” slightly more so in the treatment group. The remaining rows show potential downsides—distraction, wasting time, confusion or irrelevance, and stress—where large majorities in both roles and conditions report “not at all” or “a little.” Across all outcomes, negative impacts are rare, and patterns are similar for treatment and control, suggesting that post-meeting surveys are generally perceived as helpful with minimal cost.}
  \label{fig:posteff_postneg}
\end{figure*}

One potential reason for the effectiveness of the post-meeting surveys in the control group is that they served as a regular reminder for participants to pause for reflection over the two-week period, as noted by participants \pid{(C147, C350, C323, C169)}. The regular occurrence of such a pause may be more important than its relative timing before or after meetings. Both treatment and control group perceived the post-meeting surveys as comparably helpful for increasing awareness (\autoref{fig:posteff_postneg}).\footnote{There were no significant differences between groups on any perceptions of post-meeting surveys in \autoref{fig:posteff_postneg}.} Participants' comments reveal the initial development of a reflective habit: they \iquote{``started to think more''} \pid{(T115)}, \iquote{`became more aware''} \pid{(T471)} or \iquote{``started to notice''} \pid{(C275)} aspects of their meetings as the study progressed \pid{(C68, C485, C141)}. Moreover, treatment participants rated the post-meeting surveys as less distracting than pre-meeting surveys, particularly for organisers, suggesting that their timing may have been more suitable for reflection (comparing \autoref{fig:posteff_postneg} and \autoref{fig:preeff_posneg}: 53\% of participants rated post-meeting surveys as `not at all' distracting, compared to 39\% for pre-meeting surveys). %

Beyond reflecting in the moment \textit{after} meetings, control group participants noted how post-meeting surveys also triggered reflection \textit{before} meetings or other points in the meeting life-cycle \pid{(C68, C169, C154, C298)}. As discussed below, one driver of this may be the apparent potency of \textit{goals} (and the related concept of \textit{effectiveness}) as a reflective lens in collaborative knowledge work. 

\begin{smallquote}
    \iquote{``I feel that as the study progressed I was more inclined to think about meeting goals/agenda etc. before and after.''} \pid{(C68)}
\end{smallquote}

\subsubsection{Goals as a Meaningful Reflective Lens}\label{subsubsec:comparable-finding-reflectivelens}

Beyond being a reminder to `pause', the pre- and post- meeting surveys both focused on the intertwined concepts of meeting goals and effectiveness. Comments by control group participants illustrate that their experience with the post-meeting surveys went beyond merely evaluating the effectiveness of the preceding meeting, and was often comparable to that in the treatment group in its reflective depth. `Goals' therefore acted as a meaningful lens for stimulating reflection around meetings \textit{regardless of when or how participants were prompted}.

In both groups, reflections particularly focused on participants' \textit{roles} in meetings, including their \iquote{``impact and contributions''} \pid{(C68, C44, C77, C147, C163, C216, C427)}, what they wanted to achieve \pid{(T35, T155, T558)}, and \iquote{``whether [they] could add value''} \pid{(T115)}. Likewise, others noted that they became more aware of which meetings they were rescheduling or skipping, and why \pid{(C240, C89, C119, T454)}. Participants also noted the perspective of \textit{others} at their meetings, including how goal clarity (or lack thereof) may impact their experience \pid{(C275, C276, C517, T365)}, and their relevance to the meeting \pid{(C147, C191)}. 

\begin{smallquote}
    \iquote{``I started to notice that for meetings I organised usually goal is clear for me, but I don't make sure it's clear for everyone else.''} \pid{(C275)}
\end{smallquote}

Alongside reflecting on the \textit{in}effectiveness of many of their meetings \pid{(C27, C302, C26, C112)}, especially recurring meetings such as one-to-ones or stand-ups \pid{(C55, C191, C122, C219)}, control group participants also noticed ineffective meeting behaviours, such as conversations being dominated by people \pid{(C121)} or straying from the meeting purpose \pid{(C169, C217)}. For some, post-meeting surveys triggered reflection on \iquote{``longer term goals...  Compared to the daily grind.''} \pid{(C513)}.

The value of goals and effectiveness as a reflective lens for both groups was also observed in the actions they reported taking, which show how goals acted as a connecting thread across workflows. 

\subsubsection{Goals as an Actionable Thread Across Workflows}\label{subsubsec:comparable-finding-connectingthread}

Alongside experiencing changes in awareness, both treatment and control group participants reported taking goal-oriented actions throughout the meeting life-cycle---i.e., across time points and interfaces \textit{beyond the surveys themselves}. To this end, participants also highlighted how goals can serve as actionable inputs into AI systems for further support. Goals therefore appeared to function for participants as an \textit{actionable thread connecting across workflows}, possibly explaining why the post-meeting surveys may have been sufficient to trigger wider self-reported behaviour change.   

Among both control \pid{(C380, C298, C131, C197, C415, C427, C435)} and treatment participants \pid{(T34, T301, T207, T11, T16)}, actions included setting and reviewing meeting goals, making notes, and reflecting on how the structure of the meeting would help achieve their goals. %

\begin{smallquote}
    \iquote{``this support has been fantastic, aiding me in defining my objectives and setting expectations both before and after attending meetings.''} \pid{(C415)}
\end{smallquote}

 Better preparation also resulted in changes in pre-meeting communication. %
 For organisers in control \pid{(C37, C61, C74, C159, C188, C276, C367, C416, C419, C517)} and treatment groups \pid{(T172, T319, T414,T482, T283, T206)}, this included providing meeting goals in invitations and distinguishing these from agendas describing the meeting flow, clarifying goals at the start of meetings, or identifying the optimal times and places for sharing this information with attendees. Attendees reported clarifying meeting goals with organisers when necessary \pid{(C281, T283, T418)}. 

\begin{smallquote}
    \iquote{``Yes, i'm now alot more eager to clearly list the meeting objectives and agenda even if in bullets points for the other attendees beyond just assuming that the agenda is clear from the meeting title.''} \pid{(C276)}

\end{smallquote}

Changes were also reported as occurring during meetings. This included verbalising the goals during meetings \pid{(T418)}, discussing them with other attendees \pid{(T248, T207)}, and being \iquote{``much more conscious of not wasting time in meetings''} \pid{(T170)}. Notably, some control participants also reported such changes, e.g. being more attentive during meetings as a result of the post-meeting surveys \pid{(C141, C216, C425, C468)}. Participants also noted how goals can serve as actionable inputs into AI meeting moderation systems \pid{(C68, T311, T548)}. 

\begin{smallquote}
    \iquote{``I certainly started to consider meeting agendas and goals more consciously. [...] I was more cognizant of being an active agenda engager rather than a passive meeting attendee.''} \pid{(C216)}
\end{smallquote}

Both treatment and control group participants also noted changes in post-meeting behaviour, such as documenting take-aways \pid{(C246, C416)}, and following up on action items \pid{(C141, T34)} or goals that weren't fully met \pid{(T11)}. To this end, some participants suggested that AI systems can evaluate effectiveness with less bias than humans, using goals as benchmarks \pid{(C68, T311, T100)}, and provide learning feedback for organisers \pid{(C42, C280, T504)} and kickstart follow-up actions \pid{(C68, T32)}.

\begin{smallquote}
    \iquote{``I started thinking and solidifying my understanding of the meeting I was in and instead of just trying to remember the goals I started using copilot for every meeting to track what was said and added that to a DevOps Board which allows me to give better service to our customers.''} \pid{(C246)}

\end{smallquote}

Lastly, participants reported reviewing the meetings on their calendars for meeting goals \pid{(T256)}, their personal relevance to each meeting \pid{(T66, T192)}, and its potential value to them \pid{(T521, T379)}. Accordingly, they either declined \pid{(T256, T66, T521)}, de-prioritised \pid{(T172)}, or shifted scheduled meetings to either text chats \pid{(C199, C500)} or ad hoc calls when appropriate \pid{(C500)}.

\section{Discussion}\label{sec:discussion} 
We tested whether prompting workers to reflect on goals for upcoming meetings would improve meeting outcomes. Contrary to our hypothesis, we did not find that \textit{pre-meeting} goal reflection increased self-reported meeting effectiveness, either at the meeting level or overall. Treatment participants were also \textit{less} interested in future reflection support. This suggests that key aspects of pre-meeting goal reflection, and workplace reflection interventions more broadly, require rethinking (§\ref{designingsupport}). 

However, \textit{both treatment and control groups} showed pre-post improvements in self-reported goal clarity and communication---upstream factors that can support effective meetings---suggesting that the post-meeting surveys may have triggered comparable changes in the control group. Mixed-methods findings corroborated this and offered explanations as to why pre-meeting surveys were less impactful than expected, while post-meeting surveys were more impactful than expected. We posit that two weeks of repeatedly focusing on meeting goals may have functioned similarly to a meeting training program \cite{rogelbergSurprisingScienceMeetings2019}. This resemblance suggests that lightweight, workflow‑embedded nudges might function as forms of micro‑training, cultivating habits of intentionality and offering a foundation for designing technologies that embed reflective practices into everyday collaboration.

Our interpretation is constrained by several evaluation limitations, including issues of causal inference, reliance on self-report, intervention and study compliance, and the challenge of assessing reflection quality. We next discuss these limitations, followed by design implications.

\subsection{Evaluation Considerations and Limitations}\label{disc:eval}

\subsubsection{Causal Inference Limitations}
We interpret our null treatment effect as arising partly from %
the post-meeting surveys unintentionally functioning as a reflection intervention that also impacted the control group. This is evidenced by the quantitative analysis of pre-post changes from onboarding to debriefing (§\ref{subsubsec:goalclarityfindings}) and the qualitative analysis (§\ref{subsec:comparablechanges}). Indeed, many comments suggested participants were unaware of being in the control condition, and in that sense, the experimental design was effective \cite{boot_pervasive_2013}.

However, as our study lacked a truly passive control group---i.e., one not relying on self-report for measuring impact (see also §\ref{disc:meas} below)---we cannot exclude the potential influence of confounding external variables %
\cite{eden_field_2017,shadish2002experimental}. We are therefore limited in isolating the causal impact of the post-meeting surveys \cite{shadish2002experimental}, and leave this question for future research to confirm. This limitation aside, the observed changes in the control group are consistent with measurement reactivity, which we turn to below.

\subsubsection{Measurement Reactivity and Other Challenges of Self-Report}\label{disc:meas}

Observed changes in the control group align with measurement reactivity \cite{double_reactivity_2019}. What was intended for measurement thus potentially became an active reflection prompt, reducing our ability to detect significant group differences and therefore test the impact of our intervention. Self‑report methods such as the Experience Sampling Method ~\cite{eisele_effects_2022} or even in‑the‑moment behaviour measurement \cite{konig_systematic_2022} can also induce measurement reactivity \cite{long_mere-measurement_2025}. The introspective nature of both self‑report and reflection makes this entanglement of measurement and intervention especially acute in reflection research, particularly in workplace settings \cite{eden_field_2017}.

Alongside measurement reactivity concerns, self-reported meeting effectiveness and behaviour change is limited by people's ability to accurately assess and report their experiences and behaviour \cite{parry_systematic_2021,eisenberg_uncovering_2019, craig_evaluating_2020}, stemming from aspects such as memory limitations \cite{gorin2001recall} and social desirability concerns \cite{arnold1981social}---towards researchers and, in the meeting context, co-workers \cite{hosseinkashi_meeting_2024}. For example, we observed that organisers rated their meetings as more effective (\autoref{fig:meetingEff}), which echoes prior findings \cite{cohen_meeting_2011}, and could reflect a bias.

To address these limitations of self-report and enable a passive control condition, one solution is to measure meeting effectiveness and reflection-related behaviour change passively and unobtrusively, using automated approaches such as AI-driven transcript analysis or telemetry around invitations or participation \cite{hosseinkashi_meeting_2024, cutler_meeting_2021, zhou_role_2021}. As per §\ref{subsubsec:comparable-finding-connectingthread}, we observed a variety of changes in participants' self-reported behaviour that could be measured this way. Although promising, automated measurement of meeting effectiveness remains a technical challenge needing validation \cite{hosseinkashi_meeting_2024}. It also carries consent and surveillance concerns \cite{andrejevic2004work,soga2021web}. This constrained our approach, given that most workplace meetings in our study included non-participating workers and external partners. Automated, privacy-preserving measurement of meeting effectiveness is an important future research challenge \cite{hosseinkashi_meeting_2024}. 

\subsubsection{Intervention and Study Compliance}
We observed challenges in intervention and study compliance, with a substantial proportion of participants completing both pre- and post- meeting surveys late, or not at all. As observed in §\ref{sec:whichmeetings}, this was partly driven by survey fatigue, and was more acute in the treatment group, who had to complete twice as many surveys overall. Pre-meeting survey non-compliance may have contributed to the null treatment effect, both because participants may have not engaged sufficiently to experience change, and because non-compliance reduces available data and statistical power to detect impact \cite{jo_statistical_2002}. Our CACE analysis sought to address this by estimating impact for meetings where participants complied with the intervention \cite{peugh_beyond_2017}, but it suffers from the same statistical power limitations \cite{jo_statistical_2002}. 

To some extent, imperfect compliance is the unavoidable reality of field experiments, particularly at work, where organisational priorities conflict with research needs \cite{eden_field_2017}. However, the timing of pre-meeting surveys and the targeting of meetings, among other issues, also affected participants' engagement. We did not observe major survey timing and targeting issues in an initial pilot. However, given the centrality of intervention compliance for impact evaluation, a key learning is to proactively assess and stress-test compliance during piloting. We also discuss participant engagement and design implications in §\ref{designingsupport} below.

\subsubsection{Evaluating Reflection Quality}

A distinct measurement challenge concerns evaluating the \textit{quality} of reflection \cite{baumer_reviewing_2014,bentvelzen_revisiting_2022}. The survey method used for the pre-meeting reflection, while adapted to busy workflows, precluded elaborate externalisation that could serve as qualitative data for evaluation, particularly to better disentangle distinct impacts of the pre- and post- meeting surveys. Our analysis of open-text debriefing survey data works towards this, but is limited by the delayed, retrospective nature of participants' recollections. Future interventions using AI-assisted conversational reflection could offer richer insights both for participants and researchers \cite{scott_what_2025}.

\subsection{Designing Support for Meeting Goal Reflection Across Meetings and Workflows}
\label{designingsupport}

Our findings offer useful insights for the design of future interfaces aiming to support reflection and meeting intentionality (summarised in \autoref{tab:designimplications}).

\renewcommand{\arraystretch}{1.3}
\begin{table*}[h]
\small
\centering
\begin{tabular}{p{0.24\textwidth} p{0.01\textwidth} p{0.67\textwidth}}
\toprule
\textbf{Design Implication} && \textbf{Key Considerations} \\
\midrule
\textbf{Support Habitual Goal Reflection Without Overload} &&
\begin{tabular}[t]{@{}l@{}} 
Balance frequency to build habits without fatigue.\\
Target reflection for novel, high‑stakes, or ambiguous meetings.\\
Use contextual signals and informed consent to enable agency and avoid surveillance concerns.
\end{tabular} \\
\hline
\textbf{Embed Goal Reflection into the Rhythms and Spaces of Workflows} &&
\begin{tabular}[t]{@{}l@{}} 
Find timing ``sweet spots'' aligned with work rhythms and roles.\\
Embed reflection into scheduling tools and work canvases.\\
Match depth to context: checklists for routine, AI coaches for complex cases.\\
Enable batched reflection aligned with planning rhythms.
\end{tabular} \\
\hline
\textbf{Reinforce Goal Reflection by Making it Actionable} &&
\begin{tabular}[t]{@{}l@{}} 
Turn articulated goals into actionable artifacts across the meeting lifecycle.\\
Provide immediate returns: summaries, pre‑reads, polls, attendance optimisation etc.\\
Feed reflections into AI systems to improve planning and facilitation.\\
Link recaps and follow‑ups to evolving goals, creating a “chain of intentionality”.
\end{tabular} \\
\hline
\textbf{Support Reciprocal Goal Reflection to Drive Co-operation} &&
\begin{tabular}[t]{@{}l@{}} 
Provide shared time for reflection by blending synchronous and asynchronous modes.\\
Counteract corrosive loops with reciprocal visibility of engagement.\\
Provide evidence to challenge assumptions of disinterest or poor communication.\\
Seed virtuous cycles via habits, rhythms, and visibility. 
\end{tabular} \\
\bottomrule
\end{tabular}
\caption{Design implications for supporting goal reflection across the meeting life-cycle.}
\label{tab:designimplications}
\end{table*}
\renewcommand{\arraystretch}{1.3}

\subsubsection{Support Habitual Goal Reflection Without Overload} \label{implication-habitual}

Effective reflection systems must both prompt frequently enough to build habits, while also avoiding fatigue and desensitisation, a design tension inherent to all reflection systems. As noted in §\ref{subsubsec:comparable-finding-habitual}, participants reported survey fatigue when prompted for every meeting, yet many participants in both treatment and control groups reported increased development of reflective habits around goal clarity and communication over time.

One core aspect in the workplace context is targeting the meetings which warrant reflection. Participants questioned the value of reflecting on townhalls, multiple instances of a recurring series, or meetings with well-understood implicit goals (§\ref{sec:whichmeetings}). By contrast, meetings involving novelty, high stakes, or ambiguity are more likely to benefit from reflection \cite{scott_what_2025}. 

AI systems could target meetings based on meaningful contextual signals---such as participant diversity, project phase, or meeting history---rather than heuristics like size, length, or recurrence. However, habit formation also requires positive reinforcement. AI-driven approaches could analyse meeting characteristics to surface high-value opportunities for habit-building reflection: for example, meetings occurring during organisational change, involving ambiguous objectives, or showing declining effectiveness ratings. Routine check-ins with stable teams, by contrast, may need only periodic review.

Regardless of technology, teams must buy into a change-management process, anticipating a heightened but finite period of reflection to build habits, supported by informed consent to address surveillance concerns. A layered approach can help: agreed-upon aggregate signals of meeting effectiveness can detect systemic drift, then deeper interventions can engage only when teams see value. The goal is a culture where reflection is expected, bounded, and trusted, enabling AI systems to augment rather than erode agency \cite{tankelevitch_understanding_2025}.

\subsubsection{Embed Goal Reflection into the Rhythms and Spaces of Workflows} \label{implication-rhythms}

Goal reflection systems should find `sweet spots' in timing, placement, and interaction depth that account for the dynamic nature of workflows and priorities while enabling meaningful reflection. Our intervention five minutes before meetings proved insufficient for many participants. Yet timing too far in advance risks anticipated goals becoming irrelevant as priorities shift. This temporal tension requires role-differentiated approaches for organisers and attendees, aligned with natural work rhythms \cite{sun_rhythm_2023} (as noted in §\ref{subsubsec:relativetiming}).

For organisers, scheduling is a key time for goal reflection because this is when collaboration needs are first imagined (§\ref{sec:wherehow}). Calendar applications are thus the prime candidate for intervention, and carry implications for timing. Our study used checklists, enabling quick participation but discouraging personalisation for deeper reflection. For high-stakes or ambiguous situations, participants envisioned AI meeting coaches that guide reflection during meeting setup (§\ref{subsubsec:relativetiming}). \citet{scott_what_2025} describe such a system: an AI-driven Meeting Purpose Assistant (MPA) that briefly conversed with participants about the purpose, success conditions, and challenges of an upcoming meeting. Like a coach, it asked questions and responded to answers without offering recommendations, instead prompting participants to articulate their own understanding. For uptake, this `walk-through' must be perceived as no more effort than current interfaces, which rely on simple drop-down choices and (largely optional) field-filling. As we will discuss in §\ref{implication-actionable}, the success of such a system depends on positively reinforcing reflection by making it easily actionable in downstream workflows. %

Attendees valued having goals clearly communicated in invitations
, but often reported being too busy to do more than skim invitation information (§\ref{subsubsec:comparable-finding-connectingthread}). Embedding goal reflection and visibility into the most relevant work canvases---e.g., Customer Relationship Management software for sales teams, Project Management tools for product teams---could reach attendees when and where it matters most. This would both tie meeting goals and project communication to ongoing tasks, and enable better contextual inputs to goal reflection (see also §\ref{implication-actionable}).

Habits could also be built by \textit{batching} reflection to align with natural planning rhythms, such as at work boundaries at the beginning or end of the day \cite{sun_rhythm_2023} (§\ref{subsubsec:relativetiming}). For organisers, weekly or daily planning periods could first surface collaborative needs even \textit{before} formal scheduling. This could also enable organisers to detect recurring coordination challenges, and attendees to notice personal participation patterns.

\subsubsection{Reinforce Goal Reflection by Making It Actionable} \label{implication-actionable}

Participants saw goals as a value-laden, actionable thread connecting behaviours across meetings and workflows. Yet, they experienced a temporal mismatch in the value of reflection: the effort felt burdensome upfront, with benefits emerging only later through increased awareness. To bridge this gap, goal reflection should be designed around \textit{actionability} that justifies and reinforces the cognitive investment rather than functioning as isolated contemplation.

Actionability has three facets. First, goal reflections should produce visible resources that further strengthen individual and team intentionality. Second, these processes and artifacts should be embedded in the meeting lifecycle to support momentum and reduce coordination effort. Third, as generative AI shifts human work toward planning and orchestration \cite{passi_agentic_2025}, externalised reflections become structured inputs for AI collaboration systems, such as the scrum assistants in \citet{cabrero-daniel2024agilemeetingassistants}. Together, these elements create a flywheel: human reflection generates intent-rich data that improves AI assistance, which reduces the cognitive burden of maintaining intentionality and yields artifacts that make future meetings more effective, reinforcing continued reflection. Designing systems that capture goal articulation while returning immediate, concrete value can enhance human awareness, behaviour, and algorithmic support for more effective collaboration. The meeting lifecycle offers an ideal testbed.

Goal articulation \textit{during scheduling} could generate resources that clearly communicate meeting purpose to all attendees. For example, the Meeting Purpose Assistant \cite{scott_what_2025}, described in §\ref{implication-rhythms}, generated a short summary of participants' answers, which they intended to use in meeting invitations or personal talking points. Articulated goals can serve as important context for optimizing attendance (e.g., feeding into systems such as \cite{papachristou2025scheduling}) and engagement, for example, by surfacing information from pre-reads relevant to specific attendees, thereby addressing issues around the `lack of a pre-read culture' (§\ref{subsubsec:comparable-finding-connectingthread}).

\textit{During} meetings, AI facilitators primed with articulated goals could help participants stay aligned when discussions drift (§\ref{subsec:comparablechanges}) and link goals to inclusion considerations \cite{houtti_InclusiveAgentMeetingBot}. AI systems that map content and nudge participants about agendas and topics are increasingly common in research \cite{chen_echomind_2025_during, chen_meetmap_2025_during, chen_are_2025, chandrasegaran_talktraces_2019_during, alsobay_bringing_2025} and in commercial tools (e.g., Zoom AI Companion\footnote{https://www.zoom.com/en/products/ai-assistant}
, Microsoft Teams Meeting Facilitator\footnote{https://support.microsoft.com/en-us/office/facilitator-in-microsoft-teams-meetings-37657f91-39b5-40eb-9421-45141e3ce9f6}). Few, however, use explicit goal articulation as an input, relying instead on documents and real-time transcription. Given the complexity of human intent in collaborative knowledge work \cite{scott_what_2025, scott_mental_2024, lindley_building_2023}, we argue that current AI systems are limited in inferring intent; explicit human goal articulation remains a critical prerequisite \cite{passi_agentic_2025, kim_intentflow_2025, coscia_ongoal_2025}. AI may be better suited to evaluating how current agendas and conversations align with stated goals and intervening accordingly \cite{houtti_InclusiveAgentMeetingBot, chen_are_2025}.  
 
\textit{After and between} meetings, customised meeting recaps \cite{asthana_summaries_2025} could explicitly link meeting activity to stated goals to evaluate meeting effectiveness (though surveillance concerns arise, as per §\ref{disc:meas}). Participants suggested that goals could be dynamically updated with follow-up actions from previous occurrences (§\ref{sec:culture}). Capitalizing on the observed thread-like quality of goals, such systems could kickstart downstream workflows, for example, by proactively scheduling follow-up meetings and/or asynchronous collaboration. Further, systems could track how project objectives evolve across meeting instances, automatically connecting retrospective assessments from one meeting to prospective planning for the next. This would strengthen what \citet{vanukuru_designing_2025} describe as the ``chain of intentionality'' across related meetings.

\subsubsection{Support Reciprocal Reflection to Drive Co-operation} \label{implication-reciprocal}

Meeting intentionality is inherently interdependent, requiring coordinated effort between meeting organisers, attendees, and other collaborators. This is influenced by an organisation's meeting culture \cite{scott_orgmeetingculture_2023}, which likely interacts with personal mental models of meetings \cite{scott_mental_2024}, and personality factors associated with  meeting preparation \cite{niederman_meetingsandpersonality_1999} and conduct \cite{yoerger_meetingsandpersonality_2017}.\footnote{National culture \cite{sprain_cultureandmeetings_2012} also likely plays a role, which may well be in tension with organisational culture \cite{boussebaa_cultural_2020_orgvnatculttension, shenkar_national_2022_orgvnatculttension} and interact with personality issues. However, we do not have evidence for these factors in this study.} To address this interdependence, reflection support in a collaborative context should be designed to be \textit{reciprocal} wherever possible. 

We found that organisational norms around work intensity and limited meeting preparation undermined goal reflection. This lack of preparation fuels ineffective meetings, which then amplify work intensity, creating a corrosive loop. A reciprocal approach could be to introduce \textit{shared} time for goal reflection that blends synchronous and asynchronous work, promoting mutual engagement while reducing effort. CoExplorer \cite{park_coexplorer_2024}, for example, uses AI-generated invitation polls to let users asynchronously rank agenda priorities \cite{garcia_voting_2005}, focusing meetings on points of disagreement and leaving points of agreement for asynchronous follow-up. This helps teams better exploit the live synchronous A/V modality of meetings for effective collaboration.

Our study revealed another corrosive feedback loop: organisers assumed attendees would ignore goals and pre-reads, while attendees lacked clarity about meeting purpose and their role and disengaged (§\ref{subsec:lackchanges}). AI systems could counteract this cycle through \textit{reciprocal visibility} that makes engagement transparent to both parties. Mutual indicators could show whether attendees engaged with goals or pre-reads, while confirming that organisers communicated them clearly. This addresses the assumption asymmetry directly, providing positive evidence of mutual engagement \cite{prilla_supporting_2014}. Such visibility can foster shared ownership of meeting purpose and shift team norms toward greater intentionality.

Reciprocity exemplifies a broader design principle: technological support for meeting intentionality should \textit{seed virtuous cycles}. Rather than just fixing isolated communication breakdowns, it should establish feedback loops that gradually reshape how teams approach meetings and collaboration. The three prior implications function as interlocking facets of such cycles of reciprocity: cultivating habits (§\ref{implication-habitual}) that fit natural work rhythms (§\ref{implication-rhythms}), and making these practices visible and actionable (§\ref{implication-actionable}). Together, they form a self-reinforcing dynamic: habits reduce friction, rhythms embed reflection into everyday practice, and visibility validates and rewards engagement.

\section{Conclusion}\label{sec:conclusion}

Nudging workers to reflect on meeting goals, whether before or after meetings, fostered greater self-reported awareness and behaviour change, which are key ingredients for more intentional meetings. While survey-based pre-meeting reflection did not improve perceived meeting effectiveness, many participants expressed a strong interest in seeing goal reflection integrated into collaboration tools. These insights highlight promising opportunities to design more context-sensitive approaches that can better support reflection for workplace meetings. More broadly, such interventions can be designed to seed virtuous cycles, where small improvements in clarity and engagement reinforce one another and gradually reshape team culture toward stronger norms of intentionality.

\begin{acks}
We thank Sasa Junuzovic and Senja Filipi for their engineering support, all the participants for their time and effort, and the anonymous reviewers for their helpful feedback. 
\end{acks}

\bibliographystyle{ACM-Reference-Format}
\bibliography{references,references-sean}

\appendix
\section{Additional Methodological Details}
\subsection{Power calculations}\label{app:powercalcs}
Our target sample size was based on a power calculation conducted using a pilot dataset (n = 41), and assuming  80\% power (95\% CI: 79, 84; 1000 simulations) and 5\% alpha level, to detect a difference of 0.15 between treatment and control average meeting effectiveness ratings on a scale of 1-5 (corresponding to a standardised Cohen's \textit{d} effect size of 0.16, assuming a pooled standard deviation of 0.95 based on pilot data). 

We computed power using the `simr' package in R, for a linear mixed-effects regression on meeting effectiveness ratings (the primary outcome), with a random effect for participant ID, and fixed effects for treatment condition, whether the meeting is recurring or one-off, whether participant is the meeting organiser, and the number of invitees. Power was computed for a t-test on the treatment coefficient using the p-value calculated using the Satterthwaite method. For simplicity, and due to the small pilot sample, the regression excluded the full set of participant-level covariates and therefore represents a lower bound on power.

\subsection{Data Missingness and Multiple Imputation}\label{app:missingness}

For our primary outcome of meeting effectiveness ratings (at the meeting level), 38\% of the data was missing. To understand the nature of the missing data, we ran a mixed-effects logistic regression predicting missingness (for details see Appendix \ref{app:missingness}). We found that treatment condition, work area, whether the participant organised the meeting, whether it was a recurring meeting, and the scheduled attendee count and duration all predicted missingness (\autoref{tab:missingness}). We therefore treated the missing data as Missing at Random (MAR) conditioned on these variables and conducted multiple imputation (MI) analysis. 

\begin{table}
\footnotesize
  \caption{Results from mixed-effects logistic regression model predicting missingness in the meeting effectiveness ratings from the post-meeting surveys. `Meeting organiser', `Recurring meeting', `Scheduled attendee count', and `Scheduled duration' are meeting-level fixed effects; all others are at the participant level. Participant ID was modeled as a random effect. For the work area variable, `customer support' is used as the reference category. Statistical significance indicators: ** p < 0.001, * p < 0.01, + p < 0.1.}
  \label{tab:missingness}
  \begin{tabular}{ll}
    \toprule
    Term & Coefficient [95\% CI], p-value\\
    \midrule
TREATMENT & 0.63 [0.25, 1.02], 0.00 **\\
Recurring meeting & 0.20 [0.06, 0.34], 0.00 **\\
Meeting organizer & -0.19 [-0.33, -0.04], 0.01 *\\
Scheduled attendee count (z-scored) & 0.26 [0.19, 0.34], 0.00 ***\\
Scheduled duration (z-scored) & 0.11 [0.05, 0.18], 0.00 ***\\
Work goal specificity & -0.12 [-0.44, 0.20], 0.45\\
Overall meeting goal clarity (organizer) & -0.15 [-0.33, 0.03], 0.11\\
Overall meeting goal clarity (attendee) & 0.06 [-0.09, 0.21], 0.46\\
(Lack of) pre-meditation & 0.35 [-0.05, 0.75], 0.09 +\\
Work demands & -0.03 [-0.32, 0.25], 0.82\\
Frequency external meetings & -0.06 [-0.22, 0.11], 0.51\\
Frequency leading meetings & -0.08 [-0.30, 0.14], 0.46\\
People manager & 0.38 [-0.08, 0.83], 0.11\\
Seniority level & 0.05 [-0.24, 0.34], 0.74\\
Work area: Product development & -0.04 [-0.73, 0.64], 0.90\\
Work area: Operations & 0.07 [-0.66, 0.80], 0.86\\
Work area: Sales & 0.43 [-0.30, 1.16], 0.24\\
Work area: IT & 0.73 [-0.08, 1.55], 0.08 +\\
Work area: Other & -0.66 [-1.78, 0.46], 0.25\\
    \bottomrule
  \end{tabular}
\end{table}

We conducted MI analysis using the \texttt{mice} package, and adhered to recommendations by \citet{wijesuriya_multiple_2025}. We included all of the above variables in \autoref{tab:missingness} in our imputation model. Data was imputed using predictive mean matching, using the \texttt{2l.pmm} method for the meeting effectiveness ratings (to accommodate the hierarchical nature of the data), and \texttt{2lonly.pmm} for all participant-level variables. Meeting-level data like meeting organiser or attendee count was obtained via telemetry and did not have missing values. We followed recommendations to generate at least as many imputations as the percentage of missing cases \cite{white_multiple_2011}, i.e., 40 imputations in our case to accommodate the 38\% of missing data. We used 20 iterations to ensure convergence. We ensured the appropriateness of imputed data by examining density plots comparing imputed and raw data, and by examining traces across iterations to confirm that they overlapped and did not show trends \cite{wijesuriya_multiple_2025}. We followed a similar approach for analyses of debriefing survey measures at the participant level only; with 11\% of cases missing, we generated 20 imputations using 20 iterations.

\subsection{Regression Analysis Specifications}\label{app:regspecs}

\subsubsection{Meeting Effectiveness Ratings (Primary Outcome, Meeting Level)} 

We ran a linear mixed-effects model:
{\footnotesize
\begin{align}
\mathbf{MeetingEffectivenessRating} \sim {} & (1 \mid participantID) \nonumber \\ 
& + treatment \nonumber \\ 
& + workGoalSpecificity \nonumber \\ 
& + meetGoalClarityBaseline_{org} \nonumber \\ 
& + meetGoalClarityBaseline_{att} \nonumber \\ 
& + managerialStatus \nonumber \\ 
& + lackPreMed \nonumber \\ 
& + workDemands \nonumber \\ 
& + workArea \nonumber \\ 
& + jobLevel \nonumber \\ 
& + freqExternalMeetings \nonumber \\ 
& + freqLeadMeetings \nonumber \\ 
& + isRecurringMeeting \nonumber \\ 
& + isorganiser \nonumber \\ 
& + scheduledAttendeeCount_{zscored} \nonumber \\ 
& + scheduledDuration_{zscored}
\end{align}
}
\subsubsection{Goal Clarity and Communication (Participant Level)} 

  These outcomes were measured at both onboarding and debriefing, so we ran difference-in-differences models:
 
{\footnotesize
\begin{align}
\mathbf{Outcome} \sim {} & (1 \mid participantID) + treatment \nonumber \\
& + time + treatment \times time \nonumber \\
& + managerialStatus \nonumber \\ 
& + workGoalSpecificity \nonumber \\ 
& + lackPreMed + workDemands \nonumber \\
& + workArea + jobLevel \nonumber \\
& + freqExternalMeetings \nonumber \\
& + freqLeadMeetings
\end{align}
}

Where \textit{Outcome} is the relevant survey response across both onboarding and debriefing surveys, and \textit{time} indicates which survey. 

\subsubsection{Overall Meeting Effectiveness, Engagement, Interest in Future Support (Participant Level)} 

For these outcomes, we ran standard linear models:

{\footnotesize
\begin{align}
\mathbf{Outcome} \sim {} & treatment + workGoalSpecificity \nonumber \\
& + meetGoalClarityBaseline_{org} \nonumber \\
& + meetGoalClarityBaseline_{att} \nonumber \\
& + managerialStatus + lackPreMed \nonumber \\
& + workDemands + workArea \nonumber \\
& + jobLevel + freqExternalMeetings \nonumber \\
& + freqLeadMeetings
\end{align}
}

\subsection{GPT-4o text data labeling details}
\label{app:llmdetails}

Out of 342 responses, 22 (6\%) were manually corrected (see \autoref{tab:llmerrors}). The most common issues included instances of `awareness change' misclassified as `behaviour change' due to wording that implied \textit{potential} behaviour, but did not indicate any action was actually taken, e.g.: \iquote{``It made me realise that some meetings are repeatedly not meeting goals and to address these differently whether I'm the organiser or not.''} \pid{(C27)}. The full prompt used is provided below.

\begin{table}[h!]
\footnotesize
\caption{Contingency table comparing correct and LLM-assigned labels.}
\label{tab:llmerrors}
\centering
\begin{tabular}{c|ccc}
\multicolumn{1}{c}{} & \multicolumn{3}{c}{\textbf{Correct label}} \\ \cline{2-4}
\textbf{Assigned label} &
No change & Awareness change & behaviour change \\ \hline
No change        & 137 & 0 & 1 \\
Awareness change & 2 & 115 & 7 \\
behaviour change & 1 & 11 & 68 
\end{tabular}
\end{table}

\begin{lstlisting}[basicstyle=\footnotesize\ttfamily]
# INSTRUCTIONS
You are a survey comment analyser that reads a survey comment [body], written by a participant in a meeting goal reflection experiment, and assigns one of three labels to the comment based on the kind of change the participant experienced. The labels are:
1. "NO CHANGE" - The participant experienced no change in self-awareness or behaviour. This includes comments stating that the experience only confirmed what they already knew or did not lead to any change.
2. "SELF-AWARENESS CHANGE" - The participant experienced a change in self-awareness, such as being more aware of meetings or goals, or recognizing particular aspects of meetings or goals they hadn't considered before.
3. "BEHAVIOUR CHANGE" - The participant experienced a change in behaviour, such as improving how they approach meetings or goals, or adopting new strategies or actions related to meetings or goals.
Here are additional instructions to follow:
- Base your analysis only on the text in the survey comment.
- Avoid making assumptions or inferences beyond the explicit content of the comment.
- If the comment contains evidence for multiple labels, prioritise behaviour change over self-awareness change.
- If the comment is ambiguous or does not clearly indicate any change, assign the label "NO CHANGE."
Response:
- Return one of the three labels: "NO CHANGE", "SELF-AWARENESS CHANGE" or "BEHAVIOUR CHANGE".
- Do NOT include any other words in the output other than one of the above labels.
[body]
\end{lstlisting}

\section{Additional Findings}
\subsection{Survey Completion Times}\label{app:surveycompletiontimes}

\begin{figure}[t]
\centering
\includegraphics[width=\columnwidth]{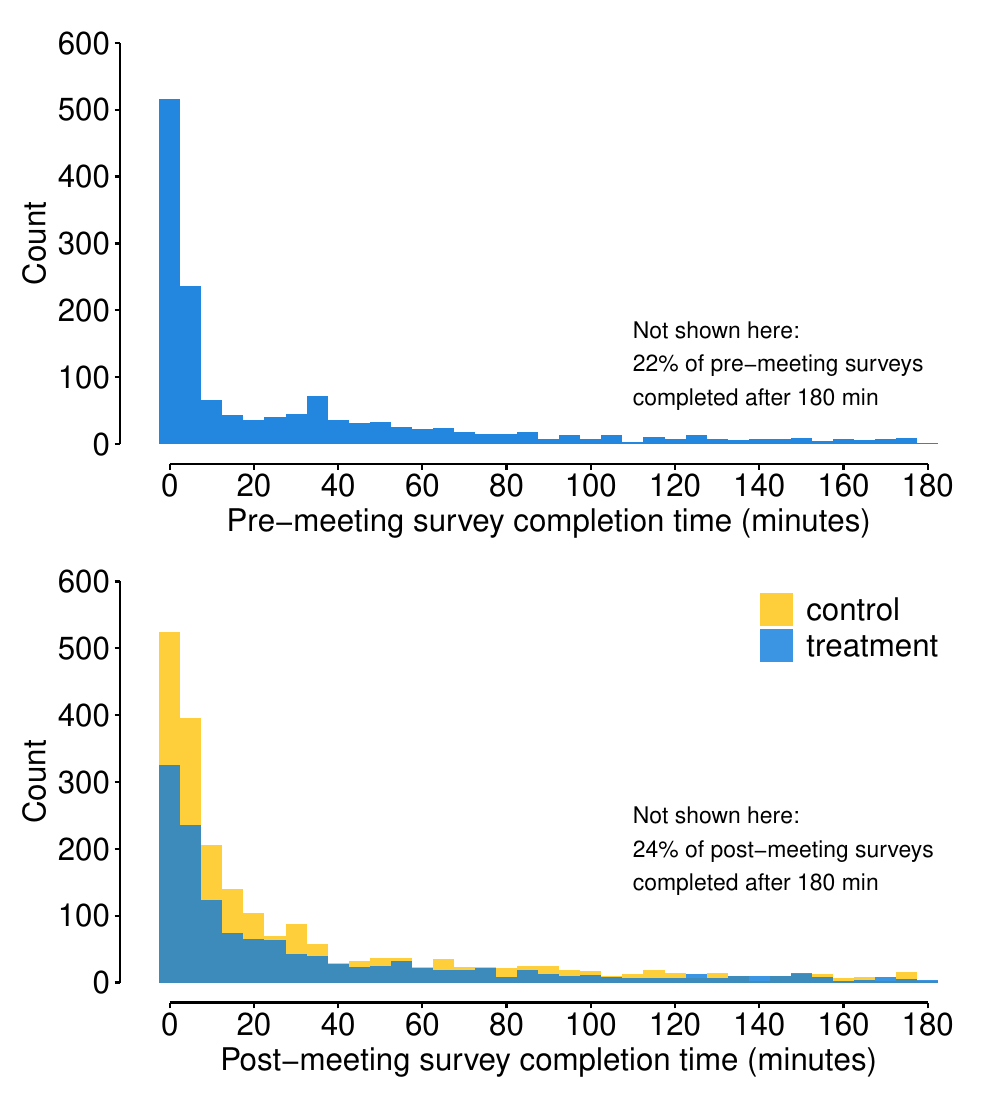}
  \caption{Distributions of survey completion times for the pre-meeting surveys (top) and the post-meeting surveys (bottom). To visualise the majority of survey completion times, the long tail of each distribution (completion times above 180 minutes) are excluded from the graphs, corresponding to 22\% and 24\% of the data, respectively.}
  \Description{Timing of pre- and post-meeting survey completion. The figure shows two histograms illustrating how long participants took to complete surveys relative to meetings. The top panel shows pre-meeting survey completion times for the treatment group, with a large spike in the first few minutes and a long right tail extending to three hours; a note indicates that 22\% of pre-meeting surveys completed after 180 minutes are not shown. The bottom panel shows post-meeting survey completion times for both treatment (blue) and control (yellow) groups, again with most responses submitted within the first few minutes and a long tail of delayed completions; 24\% of post-meeting surveys completed after 180 minutes are omitted. Overall, the figure demonstrates that most surveys are completed promptly around meetings, but a substantial minority are completed much later.}
  \label{fig:surveycompletiontimes}
\end{figure}

\autoref{fig:surveycompletiontimes} shows the distribution of pre- and post- meeting survey completion times (relative to when each survey was sent). 

\subsection{Complete Case Analysis}\label{app:completecaseresults}
To check the impact of missing data on our findings, we ran a complete case analysis of the meeting effectiveness ratings (359 participants, 4463 meetings). The pattern of findings was comparable to the MI analysis (first column in \autoref{tab:meetingEffExtras}). 

\subsection{Exact Pre-Registered Analysis}\label{app:exactprereg}
To confirm that our deviations from the pre-registration did not impact findings, we ran the primary analysis exactly as pre-registered: with original exclusion criteria (5 fewer treatment participants) and excluding `Scheduled duration' as a covariate (second column in \autoref{tab:meetingEffExtras}).   

\begin{table*}
\footnotesize
  \caption{Additional analysis results for meeting effectiveness ratings. (1) Complete case analysis (excluding any missing data), using mixed-effects linear regression model; (2) Exact pre-registered analysis, with original exclusion criteria (5 fewer treatment participants) and excluding `Scheduled duration' as a covariate, using mixed-effects linear regression model; (3) Complier Average Causal Effect (CACE) analysis, with meeting-level compliance defined as completing the pre-meeting survey within 15 minutes of delivery, using instrumental variable regression and cluster-robust standard errors to account for multi-level data. `Meeting organiser', `Recurring meeting', `Scheduled attendee count', and `Scheduled duration' are meeting-level fixed effects; all others are at the participant level. For mixed-effects models, participant ID was included as a random effect. For the work area variable, `customer support' is used as the reference category. Statistical significance indicators: *** p < 0.001, ** p < 0.01, * p < 0.05, + p < 0.1.}
  \label{tab:meetingEffExtras}
  \begin{tabular}{llll}
  
    \toprule
    Term & Coefficient [95\% CI], p-value\\
    \midrule
    & \textbf{(1) Complete case analysis} & \textbf{(2) Exact pre-registered analysis} & \textbf{(3) CACE analysis} \\
    & (359 participants, 4463 meetings) & (356 participants, 7080 meetings) & (361 participants, 7196 meetings) \\
    \midrule
TREATMENT & 0.05 [-0.09, 0.20], 0.46 & 0.06 [-0.08, 0.21], 0.40 & [-0.46, 0.77], 0.62\\
Recurring meeting & -0.28 [-0.36, -0.20], 0.00 *** & -0.28 [-0.36, -0.21], 0.00 *** & -0.25 [-0.35, -0.14], 0.00 ***\\
Meeting organizer & 0.33 [0.25, 0.42], 0.00 *** & 0.32 [0.23, 0.40], 0.00 *** & 0.34 [0.25, 0.44], 0.00 ***\\
Scheduled attendee count (z-scored) & -0.04 [-0.10, 0.01], 0.09 + & -0.03 [-0.07, 0.02], 0.25 & -0.03 [-0.09, 0.03], 0.38\\
Scheduled duration (z-scored) & 0.05 [0.01, 0.09], 0.02 * &  & 0.06 [0.01, 0.10], 0.01 *\\
Work goal specificity & 0.14 [0.03, 0.26], 0.02 * & 0.14 [0.02, 0.26], 0.02 * & 0.16 [0.02, 0.29], 0.03 *\\
Overall meeting goal clarity (organizer) & -0.01 [-0.08, 0.06], 0.82 & -0.02 [-0.10, 0.07], 0.66 & 0.00 [-0.08, 0.09], 0.95\\
Overall meeting goal clarity (attendee) & 0.15 [0.09, 0.20], 0.00 *** & 0.14 [0.08, 0.20], 0.00 *** & 0.13 [0.07, 0.19], 0.00 ***\\
(Lack of) pre-meditation & 0.02 [-0.12, 0.17], 0.75 & 0.02 [-0.13, 0.16], 0.81 & 0.03 [-0.11, 0.17], 0.69\\
Work demands & -0.12 [-0.22, -0.02], 0.02 * & -0.12 [-0.22, -0.02], 0.02 * & -0.13 [-0.24, -0.02], 0.02 *\\
Frequency external meetings & -0.10 [-0.16, -0.03], 0.00 ** & -0.09 [-0.15, -0.02], 0.01 ** & -0.08 [-0.15, -0.01], 0.02 *\\
Frequency leading meetings & 0.06 [-0.02, 0.14], 0.12 & 0.07 [-0.02, 0.15], 0.11 & 0.08 [-0.01, 0.17], 0.07 +\\
People manager & -0.05 [-0.21, 0.12], 0.59 & -0.06 [-0.23, 0.11], 0.51 & -0.05 [-0.23, 0.13], 0.57\\
Seniority level & -0.06 [-0.17, 0.05], 0.29 & -0.06 [-0.17, 0.05], 0.27 & -0.05 [-0.16, 0.07], 0.43\\
Work area: Product development & -0.33 [-0.58, -0.08], 0.01 * & -0.32 [-0.57, -0.06], 0.01 * & -0.33 [-0.60, -0.07], 0.01 *\\
Work area: Operations & -0.15 [-0.42, 0.11], 0.25 & -0.15 [-0.43, 0.12], 0.28 & -0.15 [-0.43, 0.14], 0.32\\
Work area: Sales & -0.05 [-0.31, 0.22], 0.74 & -0.06 [-0.33, 0.20], 0.63 & -0.06 [-0.34, 0.21], 0.66\\
Work area: IT & -0.22 [-0.52, 0.09], 0.16 & -0.24 [-0.55, 0.08], 0.15 & -0.20 [-0.56, 0.15], 0.25\\
Work area: Other & -0.20 [-0.60, 0.20], 0.33 & -0.18 [-0.60, 0.23], 0.39 & -0.06 [-0.48, 0.36], 0.77\\

    \bottomrule
    
  \end{tabular}
\end{table*}

\subsection{Complier Average Causal Effect (CACE) Analysis}\label{app:caceresults}

To estimate the intervention impact for meetings where participants complied with the pre-meeting survey intervention, we conducted an exploratory Complier Average Causal Effect (CACE) analysis (for results, see the third column in \autoref{tab:meetingEffExtras}). We defined compliance for treatment participants at the meeting level as completing the pre-meeting survey within 15 minutes of delivery (i.e., not completing the survey or doing so after 15 minutes was non-compliance). This resulted in 146 complying participants in the treatment group, each completing an average of 6 pre-meeting surveys across the two weeks. For control participants, compliance was always set to \texttt{FALSE} as they never had access to the pre-meeting surveys. We used random assignment to the treatment condition as the instrumental variable for compliance in an instrumental-variables regression \cite{peugh_beyond_2017} via the \texttt{AER} package. To account for the hierarchical nature of the meeting data, we ran a standard linear model and then computed cluster-robust standard errors using the \texttt{clubSandwich} package, with participant ID as the clustering variable \cite{pustejovsky_small-sample_2018}. We ran analyses on the imputed dataset and pooled results using Rubin's rules. The main instrumental variable regression specification is:

{\footnotesize
\begin{align}
\mathbf{MeetingEffectivenessRating} \sim {} & PreSurveyCompliance \nonumber \\
& + workGoalSpecificity \nonumber \\
& + meetGoalClarityBaseline_{org} \nonumber \\
& + meetGoalClarityBaseline_{att} \nonumber \\
& + managerialStatus + lackPreMed \nonumber \\
& + workDemands + workArea \nonumber \\
& + jobLevel + freqExternalMeetings \nonumber \\
& + freqLeadMeetings \nonumber \\
& + isRecurringMeeting + isOrganiser \nonumber \\
& + scheduledAttendeeCount_{zscored} \nonumber \\
& + scheduledDuration_{zscored}
\end{align}
}

The corresponding first stage is:

{\footnotesize
\begin{align}
PreSurveyCompliance \sim {} & treatment \nonumber \\
& + workGoalSpecificity \nonumber \\
& + meetGoalClarityBaseline_{org} \nonumber \\
& + meetGoalClarityBaseline_{att} \nonumber \\
& + managerialStatus + lackPreMed \nonumber \\
& + workDemands + workArea \nonumber \\
& + jobLevel + freqExternalMeetings \nonumber \\
& + freqLeadMeetings \nonumber \\
& + isRecurringMeeting + isOrganiser \nonumber \\
& + scheduledAttendeeCount_{zscored} \nonumber \\
& + scheduledDuration_{zscored}
\end{align}
}

\subsection{Interpretation of Other Correlates of Meeting Effectiveness Ratings}\label{app:meetingEffExtras}

Alongside the treatment effect, \autoref{fig:meetingEff} reports other correlates of self-reported meeting effectiveness. We observe that recurring meetings are rated as having a \textit{lower} effectiveness. Participants with high self-reported work demands, a high frequency of external meetings, and those working in product development (relative to customer support as the reference category) also tend to rate meetings as being less effective. In contrast, meetings organised by the participants and those with a longer scheduled duration are rated as having a \textit{higher} effectiveness. We also find that participants with a higher self-reported work goal specificity, and with a higher self-reported clarity of their goals (as attendees), also rate meetings as being more effective. 

\subsection{Overall Meeting Effectiveness and Engagement: Full Analysis Results}\label{app:overallmeetingeffeng}

 \autoref{tab:overall_eff_eng} shows the full results for both outcomes. 

\begin{table*}
\footnotesize
\caption{Linear regression analysis results for 
\textit{overall meeting effectiveness} and \textit{engagement}. For the work area variable, `customer support' is used as the reference category. Statistical significance indicators: *** p < 0.001, ** p < 0.01, * p < 0.05, + p < 0.1.}
\label{tab:overall_eff_eng}
\begin{tabular}{lll}
\toprule
Term & Coefficient [95\% CI], p-value & Coefficient [95\% CI], p-value\\
 & Overall Meeting Effectiveness & Overall Meeting Engagement \\
\midrule
TREATMENT & 0.10 [-0.03, 0.23], 0.13 & 0.13 [-0.00, 0.25], 0.05 +\\
Work goal specificity & -0.03 [-0.14, 0.08], 0.61 & -0.10 [-0.20, 0.01], 0.06 +\\
Overall meeting goal clarity (organizer) & -0.04 [-0.10, 0.02], 0.24 & -0.03 [-0.09, 0.03], 0.31\\
Overall meeting goal clarity (attendee) & 0.01 [-0.04, 0.06], 0.67 & 0.01 [-0.04, 0.06], 0.73\\
(Lack of) pre-meditation & -0.12 [-0.25, 0.02], 0.09 + & -0.19 [-0.32, -0.06], 0.01 **\\
Work demands & 0.04 [-0.06, 0.14], 0.43 & -0.01 [-0.10, 0.09], 0.87\\
Frequency external meetings & 0.04 [-0.02, 0.10], 0.16 & 0.01 [-0.05, 0.06], 0.77\\
Frequency leading meetings & 0.08 [0.01, 0.16], 0.03 * & 0.03 [-0.04, 0.11], 0.39\\
People manager & -0.12 [-0.27, 0.04], 0.15 & -0.04 [-0.19, 0.12], 0.65\\
Seniority level & -0.00 [-0.10, 0.10], 1.00 & -0.05 [-0.14, 0.05], 0.31\\
Work area: Product development & -0.19 [-0.43, 0.04], 0.11 & -0.13 [-0.36, 0.09], 0.25\\
Work area: Operations & -0.07 [-0.32, 0.19], 0.60 & 0.19 [-0.06, 0.43], 0.13\\
Work area: Sales & -0.13 [-0.38, 0.13], 0.33 & 0.04 [-0.20, 0.28], 0.75\\
Work area: IT & -0.05 [-0.33, 0.23], 0.73 & 0.09 [-0.18, 0.36], 0.50\\
Work area: Other & -0.21 [-0.58, 0.17], 0.28 & -0.09 [-0.45, 0.27], 0.61\\
\bottomrule
\end{tabular}
\end{table*}

\subsection{Self-Reported Goal Clarity and Communication: Full Analysis Results}\label{app:goalcomm}
\autoref{tab:goalcommDID} shows results for the difference-in-difference analyses of self-reported goal clarity and communication. For all four outcomes, there was no significant interaction between treatment and time, but there was a significant main effect of time, with participants in both groups reporting improved goal clarity and communication after the study.

\begin{table*}
\footnotesize
  \caption{Difference-in-difference linear regression models on self-reported goal clarity and communication (as organiser and attendee), which was measured before (onboarding) and after the study (debriefing). Participant ID was modeled as a random effect. For the work area variable, `customer support' is used as the reference category. Statistical significance indicators: *** p < 0.001, ** p < 0.01, * p < 0.05, + p < 0.1.}
  \label{tab:goalcommDID}
  \begin{tabular}{lllll}
\toprule
 Term & Coefficient [95\% CI], p-value &  &\\
\midrule
 & \textbf{Goal clarity (organiser)} & \textbf{Goal clarity (attendee)} & \textbf{Goal comm. (organiser)} & \textbf{Goal comm. (attendee)}\\
\midrule
TREATMENT x TIME (Debriefing) & 0.02 [-0.22, 0.26], 0.87 & -0.01 [-0.27, 0.24], 0.92 & -0.05 [-0.33, 0.24], 0.75 & -0.13 [-0.47, 0.21], 0.45\\
TREATMENT & 0.03 [-0.20, 0.25], 0.81 & 0.15 [-0.13, 0.43], 0.29 & 0.02 [-0.25, 0.29], 0.89 & -0.02 [-0.33, 0.30], 0.92\\
TIME (Debriefing) & 0.18 [0.02, 0.34], 0.03 * & 0.26 [0.08, 0.43], 0.00 ** & 0.47 [0.27, 0.66], 0.00 *** & 0.28 [0.05, 0.51], 0.02 *\\
Work goal specificity & 0.30 [0.14, 0.45], 0.00 *** & 0.31 [0.11, 0.52], 0.00 ** & 0.40 [0.21, 0.59], 0.00 *** & 0.21 [-0.01, 0.43], 0.06 +\\
(Lack of) pre-meditation & -0.22 [-0.42, -0.02], 0.03 * & -0.11 [-0.37, 0.15], 0.39 & -0.32 [-0.56, -0.08], 0.01 * & -0.51 [-0.79, -0.24], 0.00 ***\\
Work demands & -0.16 [-0.30, -0.02], 0.02 * & -0.27 [-0.45, -0.09], 0.00 ** & -0.13 [-0.29, 0.04], 0.13 & -0.18 [-0.37, 0.01], 0.07 +\\
Frequency external meetings & 0.06 [-0.02, 0.15], 0.14 & 0.05 [-0.05, 0.16], 0.33 & 0.03 [-0.07, 0.13], 0.50 & 0.12 [0.00, 0.24], 0.05 *\\
Frequency leading meetings & 0.05 [-0.05, 0.16], 0.33 & 0.19 [0.05, 0.33], 0.01 ** & 0.09 [-0.04, 0.22], 0.20 & 0.29 [0.13, 0.44], 0.00 ***\\
People manager & -0.06 [-0.29, 0.16], 0.58 & -0.28 [-0.58, 0.02], 0.07 + & -0.07 [-0.35, 0.20], 0.61 & 0.18 [-0.14, 0.50], 0.28\\
Seniority level & -0.00 [-0.14, 0.14], 0.99 & -0.12 [-0.30, 0.07], 0.23 & 0.07 [-0.10, 0.24], 0.41 & -0.19 [-0.39, 0.01], 0.06 +\\
Work area: Product development & 0.11 [-0.23, 0.45], 0.53 & -0.19 [-0.63, 0.26], 0.41 & 0.05 [-0.36, 0.46], 0.82 & -0.14 [-0.61, 0.34], 0.57\\
Work area: Operations & 0.28 [-0.08, 0.65], 0.13 & 0.09 [-0.39, 0.57], 0.72 & 0.23 [-0.21, 0.67], 0.31 & 0.14 [-0.37, 0.65], 0.59\\
Work area: Sales & 0.13 [-0.24, 0.49], 0.49 & -0.20 [-0.68, 0.27], 0.41 & 0.10 [-0.34, 0.53], 0.66 & -0.02 [-0.53, 0.49], 0.93\\
Work area: IT & 0.10 [-0.31, 0.51], 0.63 & 0.35 [-0.18, 0.89], 0.20 & 0.23 [-0.26, 0.72], 0.36 & 0.02 [-0.55, 0.59], 0.94\\
Work area: Other & 0.42 [-0.12, 0.96], 0.13 & -0.13 [-0.84, 0.59], 0.73 & 0.29 [-0.37, 0.94], 0.39 & -0.44 [-1.21, 0.32], 0.25\\
\bottomrule
  \end{tabular}
\end{table*}

\subsection{Interest in Future Reflection Support: Full Analysis Results}\label{app:interestPrePrompt}

 \autoref{tab:interestPrePost} shows the full results for both outcomes.

\begin{table*}
\footnotesize
  \caption{Linear regression analysis results for interest in future pre- and post- meeting prompts for goal reflection. For the work area variable, `customer support' is used as the reference category. Statistical significance indicators: *** p < 0.001, ** p < 0.01, * p < 0.05, + p < 0.1.}
  \label{tab:interestPrePost}
  \begin{tabular}{lll}
    \toprule
    Term & Coefficient [95\% CI], p-value & \\
    \midrule
    & Pre-Meeting Prompt & Post-Meeting Prompt\\
    \midrule
TREATMENT & -0.34 [-0.61, -0.07], 0.01 * & -0.32 [-0.59, -0.04], 0.03 *\\
Work goal specificity & 0.03 [-0.19, 0.26], 0.77 & 0.10 [-0.13, 0.33], 0.39\\
Overall meeting goal clarity (organizer) & -0.03 [-0.16, 0.10], 0.65 & 0.02 [-0.11, 0.15], 0.73\\
Overall meeting goal clarity (attendee) & -0.07 [-0.18, 0.04], 0.20 & -0.07 [-0.18, 0.03], 0.18\\
(Lack of) pre-meditation & -0.07 [-0.36, 0.21], 0.61 & -0.21 [-0.50, 0.07], 0.15\\
Work demands & -0.08 [-0.28, 0.12], 0.43 & -0.11 [-0.31, 0.10], 0.30\\
Frequency external meetings & 0.03 [-0.09, 0.15], 0.60 & 0.08 [-0.04, 0.20], 0.18\\
Frequency leading meetings & -0.03 [-0.19, 0.12], 0.68 & 0.02 [-0.14, 0.18], 0.80\\
People manager & 0.17 [-0.15, 0.50], 0.30 & 0.00 [-0.33, 0.33], 1.00\\
Seniority level & -0.02 [-0.23, 0.18], 0.81 & -0.07 [-0.27, 0.14], 0.52\\
Work area: Product development & -0.19 [-0.67, 0.29], 0.44 & -0.00 [-0.50, 0.49], 0.99\\
Work area: Operations & -0.13 [-0.65, 0.39], 0.62 & -0.07 [-0.60, 0.47], 0.81\\
Work area: Sales & 0.26 [-0.26, 0.78], 0.33 & 0.20 [-0.33, 0.72], 0.46\\
Work area: IT & 0.17 [-0.41, 0.74], 0.57 & 0.43 [-0.15, 1.02], 0.15\\
Work area: Other & -0.48 [-1.24, 0.29], 0.22 & -0.15 [-0.93, 0.63], 0.71\\
    \bottomrule
  \end{tabular}
\end{table*}

\end{document}